\documentclass[epj,amsmath]{svjour}
\usepackage{color,graphicx,xcolor,ulem,float}
\usepackage[utf8]{inputenc}

\def\roughly#1{\mathrel{\raise.3ex\hbox{$#1$\kern-.75em%
\lower1ex\hbox{$\sim$}}}}
\def\lsim{\roughly<}
\def\gsim{\roughly>}

\begin{document}
\title{Hyperons in neutron stars and supernova cores}
\subtitle{}
\author{Micaela Oertel\inst{1} \and Francesca Gulminelli\inst{2}% etc
\and 
Constança Providência\inst{3} \and Adriana R. Raduta\inst{4}% etc
}                     % Do not remove
\institute{LUTH, CNRS, Observatoire de Paris, Universite Paris Diderot, 
5 place Jules Janssen, 92195 Meudon, France\and
ENSICAEN, UMR6534, LPC ,F-14050 Caen c\'edex, France\and
CFisUC, Department of Physics, University of Coimbra, Portugal \and
IFIN-HH, Bucharest-Magurele, POB-MG6, Romania
}
\date{Received: date / Revised version: date}
\abstract{The properties of compact stars and their formation
  processes depend on many physical ingredients. The composition and
  the thermodynamics of the involved matter is one of them. We will
  investigate here uniform strongly interacting matter at densities
  and temperatures, where potentially other components than free
  nucleons appear such as hyperons, mesons or even quarks. In this
  paper we will put the emphasis on two aspects of stellar matter with
  non-nucleonic degrees of freedom. First, we will study the phase
  diagram of baryonic matter with strangeness, showing that the onset
  of hyperons, as that of quark matter, could be related to a very
  rich phase structure with a large density domain covered by phase
  coexistence. Second, we will investigate thermal effects on the
  equation of state (EoS), showing that they favor the appearance of
  non-nucleonic particles. We will finish by reviewing some recent
  results on the impact of non-nucleonic degrees freedom in compact
  star mergers and core-collapse events, where thermal effects cannot
  be neglected.
\PACS{
      {PACS-key}{discribing text of that key}   \and
      {PACS-key}{discribing text of that key}
     } % end of PACS codes
} %end of abstract
\maketitle
\section{Introduction}
\label{intro}

The properties of compact stars, their formation processes as well as
binary mergers depend on many different physical ingredients, among
them the thermodynamic properties of the involved matter entering via
the equation of state (EoS).  There is an intrinsic connection between
the properties of matter contained in the EoS for the macroscopic
description of astrophysical objects and the underlying fundamental
interactions between particles on the microscopic level.  This makes
the study of the aforementioned systems very rewarding as they
challenge our understanding of nature on both scales.

It is not an obvious task to construct such an EoS.  The main
difficulty arises from the fact that very large ranges of (baryon
number) densities ($10^{-10}\ \mathrm{fm}^{-3}\lsim n_B\lsim
1\ \mathrm{fm}^{-3}$), temperatures ($0<T\lsim150$ MeV) and hadronic
charge fractions ($0< Y_Q = n_Q/n_B \lsim\, 0.7$) have to be
covered. $n_Q$ here denotes the total hadronic charge density, which
in many cases is just given by the proton density. Within this range,
the characteristics of matter change dramatically, from an ideal gas
of different nuclei up to uniform strongly interacting matter,
containing in the simplest case just free nucleons and potentially
other components such as hyperons, nuclear resonances or mesons. Even
a transition to deconfined quark matter cannot be excluded.

For core collapse matter, the full density, temperature, and
$Y_q$-dependence have to be included within the EoS. This complexity is
the main reason why until recently only a few hadronic EoSs existed
for core collapse simulations. These are the one by Hillebrandt and
Wolff \cite{hillebrandt1984}, used by some groups performing supernova
simulations, that by Lattimer and Swesty~\cite{Lattimer:1991nc} and finally
that by H. Shen \textit{et al.}~\cite{shen_98}. The two latter,
publicly available, are most commonly used in core-collapse
simulations. They use different nuclear interactions, but are based on
the same limiting assumptions: they take into account non-interacting
$\alpha$-particles, a single heavy nucleus and free nucleons in
addition to the electron, positron and photon gas.
 
In recent years, several new models have been constructed, enlarging
the variety of nuclear interaction models. This helps to estimate the
uncertainty on astrophysical simulations induced by our limited
knowledge about the interaction in hot and dense matter. Apart from
employing different models, the full nuclear distribution at
sub-nuclear densities has been included within different approaches
(see e.g.~\cite{Blinnikov:2009vm,Heckel09,Hempel09,Shen:2011kr,shen_11_nl3,Sumiyoshi08a,Raduta10}), showing considerable differences to that
obtained via the single nucleus approximation in the standard EoS
employed in core collapse simulations~\cite{Lattimer:1991nc,shen_98}.  

Up to now, much less effort has been devoted to the high density ($n_B
\gsim n_0$) and high temperature ($T \gsim 20$ MeV) part of the EOS,
including additional particles, such as hyperons and mesons or quarks.
One reason might be that the recent observation of two neutron stars
with a mass of about $2 M_\odot$~\cite{Antoniadis_13,Demorest_10} has
triggered intensive discussion on the composition of matter in the
central part of neutron stars and its EoS, excluding in the standard
picture additional degrees of freedom in super-saturation matter, i.e. at densities above a baryon number density of $n_0 \approx 0.16$ fm$^{-3}$, since
they lead to a considerable softening of the
EoS~\cite{Glendenning:1982nc}. This observation, however, does not exclude
them, but only puts stringent constraints on the respective
interaction (e.g.~\cite{Hofmann00,RikovskaStone:2006ta,Bednarek:2011gd,Weissenborn11b,Weissenborn11c,Bonanno11,Oertel:2012qd,Colucci2013,Lopes2014,Banik2014,vanDalen2014,Katayama2014,Oertel_14}). Different solutions with hyperonic and/or quark matter
have been proposed without any definite conclusion. 

In addition, even if it turns out that finally in cold neutron stars
only nucleonic matter is present, in stellar core-collapse events and
neutron star mergers, matter is strongly heated in addition to being
compressed to densities above nuclear matter saturation density. The
temperatures and densities reached can become so high that a
traditional description in terms of electrons, nuclei, and nucleons is
no longer adequate. Compared with the cold neutron star EoS,
temperature effects favor the appearance of additional particles such
as pions and hyperons and they become abundant in this regime.  A
transition to quark matter is possible, too~\cite{sagert_09}. 

The opening of additional degrees of freedom in dense matter could
happen smoothly or could be accompanied by a phase transition with a
considerable effect on the thermodynamics and the hydrodynamical
evolution of the system, see e.g.~\cite{sagert_09}. In this context,
the best known example is the hadron-quark phase transition, expected
to be first order within the density and temperature range relevant
for compact stars and core-collapse supernovae, see e.g. the
contribution by A. Sedrakian to this
issue~\cite{Sedrakian_this_issue}. In practice, anyway, the hadronic
and the quark phases are described within different models, the
transition necessarily shows discontinuities in the thermodynamic
quantities. It is less known that the appearance of hyperons could be
associated with a ``strangeness driven'' phase transition, too,
similar to the liquid-gas transition in nuclear matter. A detailed
study of the phase diagram of the $n,p,\Lambda$-system was recently
undertaken in Refs.~\cite{Gulminelli:2012iq,Gulminelli:2013qr} within a non-relativistic
mean-field model based on phenomenological functionals. It was shown
that under these assumptions first- and second- order phase transition
exist, and are expected to be explored under the strangeness
equilibrium condition characteristic of stellar matter. In
Refs.~\cite{SchaffnerBielich:2000wj,Oertel_14} a phase transition at the onset
of hyperons has been discussed for relativistic mean field models,
however in a model with very strong $YY$ attraction.

Within this paper, we want to address two aspects of the EoS of
super-saturation matter. First, we will investigate the influence of a
possible phase transition at the respective hyperonic thresholds on
the phase diagram of super-saturation baryonic matter and discuss
possible astrophysical consequences. We will not discuss the question
whether hyperonic interactions lead to such a phase transition or not,
see e.g.~\cite{James_15} for a thorough study of that point, but just
assume that it exists and explore the possible consequences.  Second,
we will, independently of the way they appear, discuss different works
on the super-saturation EoS including additional particles with an
emphasis on thermal effects. The cold neutron star part will be
discussed in other contributions to this
issue~\cite{Chatterjee_this_issue}. We will mention astrophysical
applications, too, discussing in particular the impact of additional
particles on black hole formation.

This paper is organized as follows: in Sec. \ref{section:PT}, we
review the main features of the
thermodynamic analysis of the phase diagram  of a
$\cal N$-component system, and the phase diagram of baryonic matter
with strangeness is discussed, in particular, a possible strangeness
driven phase transition and the Coulomb effects on the phase diagram;
in Sec. \ref{sec:sneos_non_uniform_additional} thermal effects on
the EoS including non-nucleonic degrees of freedom are presented, we
review relativistic mean field models for the EoS, including the
hyperonic interaction, and discuss the effect of including hyperons and pions in the finite temperature EoS;
in Sec. \ref{sec:astro} the impact of additional particles in
astrophysical applications is referred and finally in the last section
some conclusions are drawn.

\section{A possible strangeness-driven phase transition?}
\label{section:PT}
\subsection{Thermodynamic analysis of the phase diagram}
\label{section:thermodynamics}
In this section we would like to recall the main features of the
thermodynamic analysis of the phase diagram, see
e.g.~\cite{Glendenning92}. The phase diagram of a $\cal N$-component
system is, at constant temperature, a $\cal N$-dimensional volume. If
there are $M$ different coexisting phases, the boundaries of the phase
coexistence domain(s), $\{n_i^{{\cal P}_j}\}; ~ i=1,...,{\cal N};
~j=1,...,M$, are determined by the $({\cal N}+1)(M-1)$ conditions of
thermodynamic equilibrium between them,
\begin{eqnarray}
\left( \frac{\partial f}{\partial n_i} \right)_{{\cal P}_1}=
...=\left( \frac{\partial f}{\partial n_i} \right)_{{\cal P}_M}=\mu_i; 
~~ i=1,...,{\cal N}
\nonumber \\
\left(-f + \sum_i n_i \frac{\partial f}{\partial n_i} \right)_{{\cal P}_1}=...
=\left(-f + \sum_i n_i \frac{\partial f}{\partial n_i} \right)_{{\cal P}_M}=P, \nonumber \\
\label{eq:equil}
\end{eqnarray}
where $f$ denotes the free energy density and $P$ the pressure.
Within the phase coexistence domain(s), a mixture of different phases
lowers the free energy of the system as compared with the solutions
corresponding to individual phases. Mathematically, this is equivalent
to the presence of a convexity anomaly of the thermodynamic potential
in the density hyperspace, i.e. the free energy curvature matrix,
$C_{ij}=\partial^2 f/\partial n_i\partial n_j$, has at least one
negative eigenvalue . The number of coexisting phases is determined by
the number of order parameters~\footnote{An order parameter is defined
  here as the direction in the observable space corresponding to phase
  separation, see e.g. Ref.~\cite{Chomaz06} for details} or, in terms
of local properties, by the number of directions in density space
where spinodal instabilities develop, i.e. where density fluctuations are spontaneously amplified finally leading
to phase separation.  The latter quantity is related to the
number ${\cal N}_{\mathit{neg}}$ of negative eigenvalues of $C_{ij}$,
such that $M ={\cal N}_{\mathit{neg}}+1$.

If the direction of phase separation is unique, which corresponds to a
one-dimensional order parameter, then the problem of phase coexistence
in a $\cal N$-component system can by reduced to a problem of phase
coexistence in a one-component system by Legendre transforming the
thermodynamical potential $f$ with respect to the remaining $({\cal
  N}-1)$-chemical potentials~\cite{Ducoin:2005aa}.

Baryonic matter with hyperons contains eight different particle
species.  Under the condition of equilibrium with respect to the
strong interaction, the number of relevant degrees of freedom is,
however, reduced to three, the densities baryon number density $n_B$,
the total baryonic (electric) charge density $n_Q$ and the total
strangeness density $n_S$. It represents, therefore, a three-component
system in the terminology introduced above~\footnote{See
  Ref. \cite{Gusakov} for a discussion of the phase diagram if strong
  equilibrium is not assumed.} It is important to remark here that the
use of strangeness as a relevant degree of freedom does not imply that
$n_S$ is conserved throughout the evolution of the system. In
particular, along the strangeness equilibrium trajectory $\mu_S=0$
considered in this study, $n_S$ obviously varies.  Upon adding
leptonic degrees of freedom in form of electrons and positrons, (electron) lepton
number enters as an additional variable. Due to the strict
electrical neutrality condition, however, charge is no longer an
independent degree of freedom once leptons are included and the system remains
three-dimensional~\cite{Chomaz:2005xn,Gulminelli:2013qr}, see also
Ref.~\cite{Providencia:2006mm}, in terms of the number densities $n_B,
n_S$ and $n_L$. In the absence of neutrinos, $n_L$, the electron
lepton density, denotes here the net electron density, $n_L = n_e =
n_{e^-} - n_{e^+}$ and charge neutrality gives $n_L = n_Q$

The equilibrium conditions thus reduce the dimensionality of the phase
space from 8 (9 with electrons), corresponding to the number of
different particle species, to three, described by baryon, lepton and
strangeness number densities. To further reduce the dimensionality for
studying phase coexistence, one may then perform the Legendre
transformation with respect to any set $(\mu_B,\mu_S)$,
$(\mu_S,\mu_{Q(L)})$ and $(\mu_B,\mu_{Q(L)})$. $\mu_{Q(L)}$ thereby
stands either for $\mu_Q$ in a purely baryonic systems or for $\mu_L$
if electrons are included. In practice, we found that the order
parameter is always one dimensional.  This means that a single
Legendre transformation is enough to spot the thermodynamics provided
that the order parameter is not orthogonal to the controlled density.
The most convenient framework to easily access the physical
trajectories is the one controlling the $n_B$-density:
\begin{equation}
\bar f(n_B,\mu_S,\mu_{Q(L)})=f(n_B,n_S,n_{Q(L)})-\mu_S n_S - \mu_{Q(L)} n_{Q(L)}.
\end{equation}
Coexisting phases, if any, will then be characterized by equal
values of $\mu_B=\partial \bar f/\partial n_B$ and $P$ and the phase
instability regions will be characterized by a back-bending behavior
of $\mu_B(n_B)|_{\mu_S,\mu_{Q(L)}}$.

\subsection{The phase diagram of baryonic matter with strangeness}
\label{section:uncharge}

Due to the large incompressibility of electrons present to ensure
electrical charge neutrality, Coulomb interactions can have a
considerable influence on the phase diagram, in particular if charged
particles are involved. For instance, the well known nuclear
liquid-gas phase transition is strongly quenched by Coulomb
effects~\cite{Ducoin:2005aa,Barranco:1981zz,DoHaMe2000}. On the
contrary, in Ref.~\cite{Gulminelli:2013qr} it has been shown that, if
$\Lambda$-hyperons are the only strange baryons, 
the supra-saturation phase transition
occurring in the $np\Lambda$+$e$-system is only slightly modified by
Coulomb interactions compared with the $np\Lambda$-system. The reason
is that this phase transition is ``strangeness driven'', i.e. the
order parameter is dominated by $n_S$ and has only a small component
in direction of the charge.  On the other hand, if the hyperonic
couplings are such that charged hyperons are abundant, as predicted by
many models favoring negatively charged hyperons, the extension and
localization of phase coexistence domains could be strongly modified
by the Coulomb interaction. In the extreme case, it could even make it
disappear for neutron star matter.

For the present study we will employ the
non-relativistic energy density functional developed by Balberg and
Gal~\cite{Balberg97}, which is known to present a rich phase
structure~\cite{Gulminelli:2012iq,Gulminelli:2013qr}. The total baryonic energy density is then given by the sum of mass, kinetic and potential energy density,
\begin{equation}
 e_B=\sum_{i=n,p,Y}\left ( n_i m_i c^2+ \frac{\hbar^2}{2m_i} \tau_i \right )
+ e_{pot}(\{n_i\})~. 
\end{equation}
The single-particle densities and kinetic energy densities are thereby
given by the Fermi integrals
\begin{equation}
n_i
=\frac{4 \pi }{h^3} \left(\frac{2m_i}{\beta} \right)^{\frac 32} 
F_{\frac 12}(\beta \tilde \mu_i)  ; \;
\tau_i=\frac{8\pi^3 }{h^5} \left( \frac{2 m_i}{\beta } \right )^{\frac 52}
F_{\frac 32}(\beta \tilde \mu_i),
\end{equation}
%%%%%%%%%%%%%%%%
with $F_{\nu}(\eta)=\int_0^{\infty} dx \frac{x^{\nu}}{1+\exp \left(
  x-\eta\right)}$ being the Fermi-Dirac integral, $\beta=T^{-1}$ is
the inverse temperature, $m_i$ denotes the $i$-particle mass and
$\tilde \mu_i$ the effective chemical potential of particle species $i$-species.

The potential energy density proposed by Balberg and Gal~\cite{Balberg97} has the following form,
\begin{eqnarray}
e_{\mathit{pot}}^{(BG)}\left( \{ n_i\} \right)&=&\sum_{i,j} e_{ij}^{(BG)} (n_i,n_j); \nonumber \\
e_{ij}^{(BG)} (n_i,n_j)&=&\left( 1-\frac{\delta_{ij}}{2} \right)
( a_{ij} n_i n_j +b_{ij}  n_{i3} n_{j3} \nonumber \\ 
&+&c_{ij} \frac{n_i^{\gamma_{ij}+1} n_j + n_j^{\gamma_{ij}+1} n_i}{n_i+n_j}
),
\label{eq:epot_BG}
\end{eqnarray} 
For simplicity the same functional form is employed in all
  channels and a unique value is used for the exponent govering the
  short-range repulsion.  The values of the coupling constants, listed
  in Table~\ref{tab:bg1}, have been chosen to satisfy the experimental
  constraints available at that time, symmetry arguments or, in the
  case of the $NN$-channel, to agree with popular models.  Concerning
  the latter, the resulting values are in good agreement with
  experimental contraints for the saturation density, $n_0 =$ 0.155
  fm$^{-3}$, and energy per nucleon of symmetric nuclear matter at
  saturation, -15.9 MeV.  The incompressibility modulus with 375
  MeV is, instead, largely overestimated.  The well depths of various
  hyperonic species in uniform symmetric nuclear matter at saturation
  density are: $U_{\Lambda}^{(N)}(n_0)=U_{\Sigma}^{(N)}(n_0)=-26.6$
  MeV, $U_{\Xi}^{(N)}(n_0)=-22.8$ MeV and the value of
  $\Lambda$-potential in $\Lambda$-matter
  $U_{\Lambda}^{(\Lambda)}(n_0/5)$=-12.8 MeV.  Both, the $\Xi N$- and
  $\Lambda \Lambda$-potentials are too attractive in view of actual
  experimental data.  AGS-E885 data indicate for $U_{\Xi}^{(N)}(n_0)$
  the value of -14 MeV \cite{Khaustov99}.  The value of
  $U_{\Lambda}^{(\Lambda)}(n_0/5)$ can be related to the bond energy
  of double-$\Lambda$ hypernuclei~\cite{Vidana01}. Experimental data for
  $_{\Lambda \Lambda}^6$He \cite{Ahn2013} suggest that a much larger
  value of $U_{\Lambda}^{(\Lambda)}(n_0/5)$ = -0.67 MeV is more
  realistic.  The situation of the $\Sigma N$-potential is ambigous
  but, very probably, it is repulsive.  Alternative values of the
  coupling constants which overcome the drawbacks of the original
  parameterization have been proposed in Ref. \cite{Oertel:2012qd}.

Thermodynamic consistency allows to infer the relation between the
chemical potentials $\mu_i \equiv \left. \partial e_{B}/\partial
n_i\right|_{n_j, j\neq i} $ and the effective ones $\tilde \mu_i$ as
$\mu_i=\tilde \mu_i + m_i c^2 +U_i$, with $U_i=\partial
e_{pot}/\partial n_i$. All remaining thermodynamic quantities, such as
pressure or entropy can then be derived from the energy density via
standard thermodynamic relations. 

\begin{table*}
\begin{center}
\caption{Coupling constants corresponding to the stiffest interaction 
proposed in Ref.~\cite{Balberg97}, called BGI within this paper, and those corresponding 
to the nuclear interaction used by Lattimer and Swesty~\cite{Lattimer:1991nc} 
supplemented with the $\Lambda$-hyperon~\cite{Gulminelli:2013qr,Peres_13}.
In case of BGI, the only hyperon pair with non-vanishing isospin-component is $\Sigma$-$\Sigma$.\label{tab:bg1}}
\resizebox{1.99\columnwidth}{!}{%
\begin{tabular}{ccccccccccccc}
\hline
\hline
Parameter& $a_{NN}$ & $b_{NN}$  & $c_{NN}$ &
$a_{\Lambda\Lambda}$ & $b_{\Lambda\Lambda}$ & $c_{\Lambda\Lambda}$ & $a_{\Lambda N}$ &$b_{\Lambda N}$ &
$c_{\Lambda N}$ & $\gamma_{NN}$ & $\gamma_{\Lambda N}$ & $\gamma_{\Lambda \Lambda}$ \\
set & MeV fm$^3$ & MeV fm$^3$   & MeV fm$^{3\gamma_{NN}}$  &  MeV fm$^3$  &
MeV fm$^{3}$  & MeV fm$^{3\gamma_{\Lambda \Lambda}}$ & MeV fm$^3$ & MeV fm$^3$  &  MeV fm$^{3\gamma_{\Lambda N}}$  &  &  & \\
\hline
BGI             &-784.4 &  214.2 & 1936.0 & -486.2 & 0 & 1553.6 & -340.0 & 0 &
1087.5 & 2 & 2 & 2\\
LS220$\Lambda$  &-1636.2 &  214.2 & 1869.2 & -486.2 & 0 & 1553.6 & -340.0 & 0 &
1087.5 & 1.26 & 2 & 2\\
220g3           &-1636.2 & 214.2 & 1869.2 & -90.0   & 0 & 1000.0 & -270.0 & 0 &
4000.0 & 1.26 & 3 & 3 \\
\hline
\end{tabular}
}
\resizebox{1.99\columnwidth}{!}{%
\begin{tabular}{ccccccccccccc}
\hline
Parameter& $a_{\Sigma N}$ & $b_{\Sigma N}$  & $c_{\Sigma N}$ & $a_{\Xi N}$ & $b_{\Xi N}$  & $c_{\Xi N}$ & 
$a_{YY}$ & $b_{YY}$  & $c_{YY}$ & 
$\gamma_{\Sigma N}$ & $\gamma_{\Xi N}$ & $\gamma_{YY}$\\
set & MeV fm$^3$ & MeV fm$^3$   & MeV fm$^{3\gamma_{\Sigma N}}$ & MeV fm$^3$ & MeV fm$^3$   & MeV fm$^{3\gamma_{\Xi N}}$ 
& MeV fm$^3$ & MeV fm$^3$   & MeV fm$^{3\gamma_{YY}}$ 
& & & \\
\hline
BGI             &-340.0 & 214.2 & 1087.5 & -291.5 & 0 & 932.5 & $a_{\Lambda \Lambda}$ & 0/428.4 & 
$c_{\Lambda \Lambda}$ & 2& 2 &2 \\ 
220g3 & 450.0 & 214.2 & 250.0 & -170.0 & 0.0 & 2900.0 & $a_{\Lambda \Lambda}$ & 0/430.0 & 
$c_{\Lambda \Lambda}$ & 3 & 3 &3 \\
\hline
\hline
\end{tabular}
}
\end{center}
\end{table*}

\subsubsection{The $n,p,Y$-system without electrons}
\begin{figure}
\begin{center}
\includegraphics[angle=0, width=0.99\columnwidth]{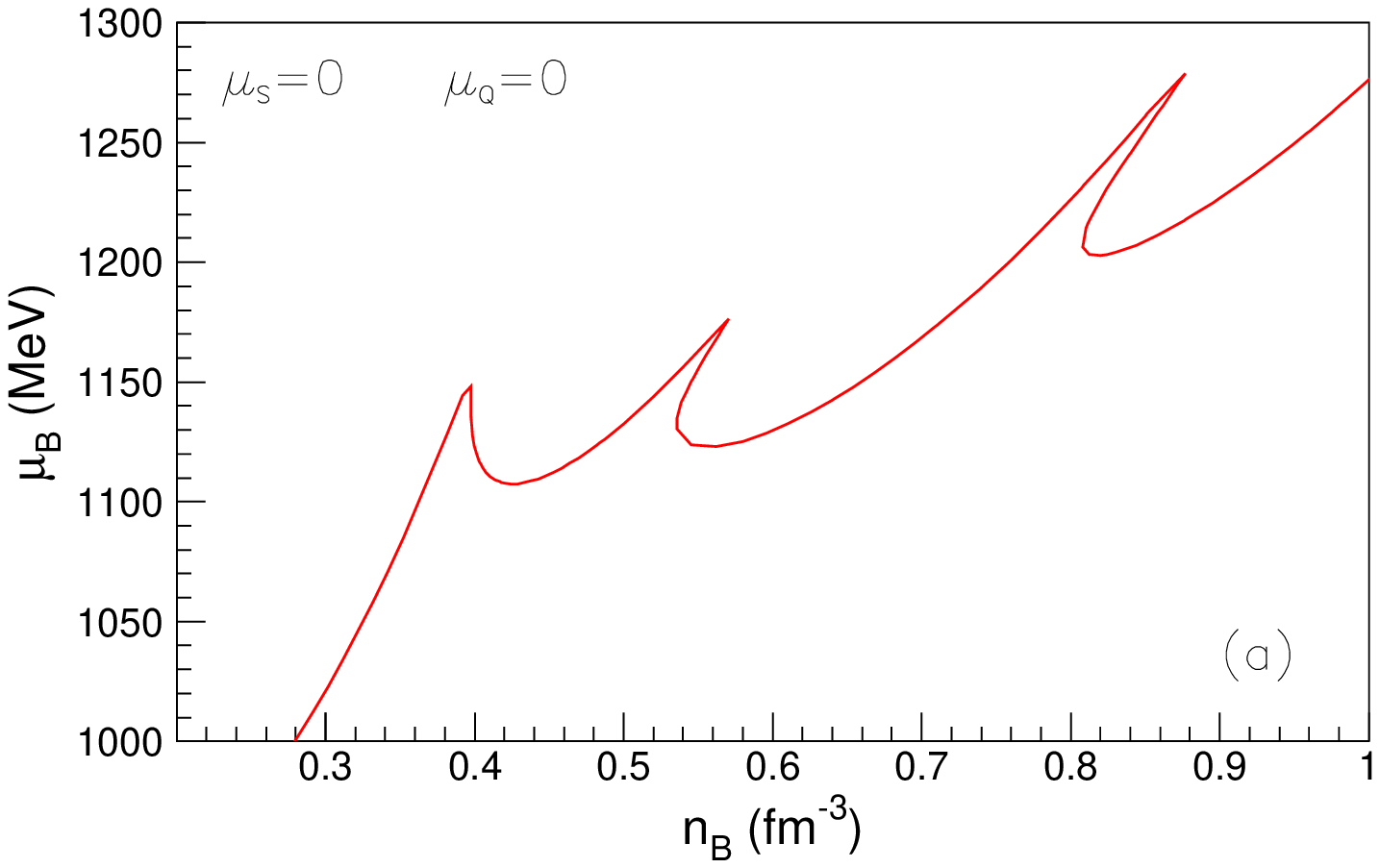}
\includegraphics[angle=0, width=0.99\columnwidth]{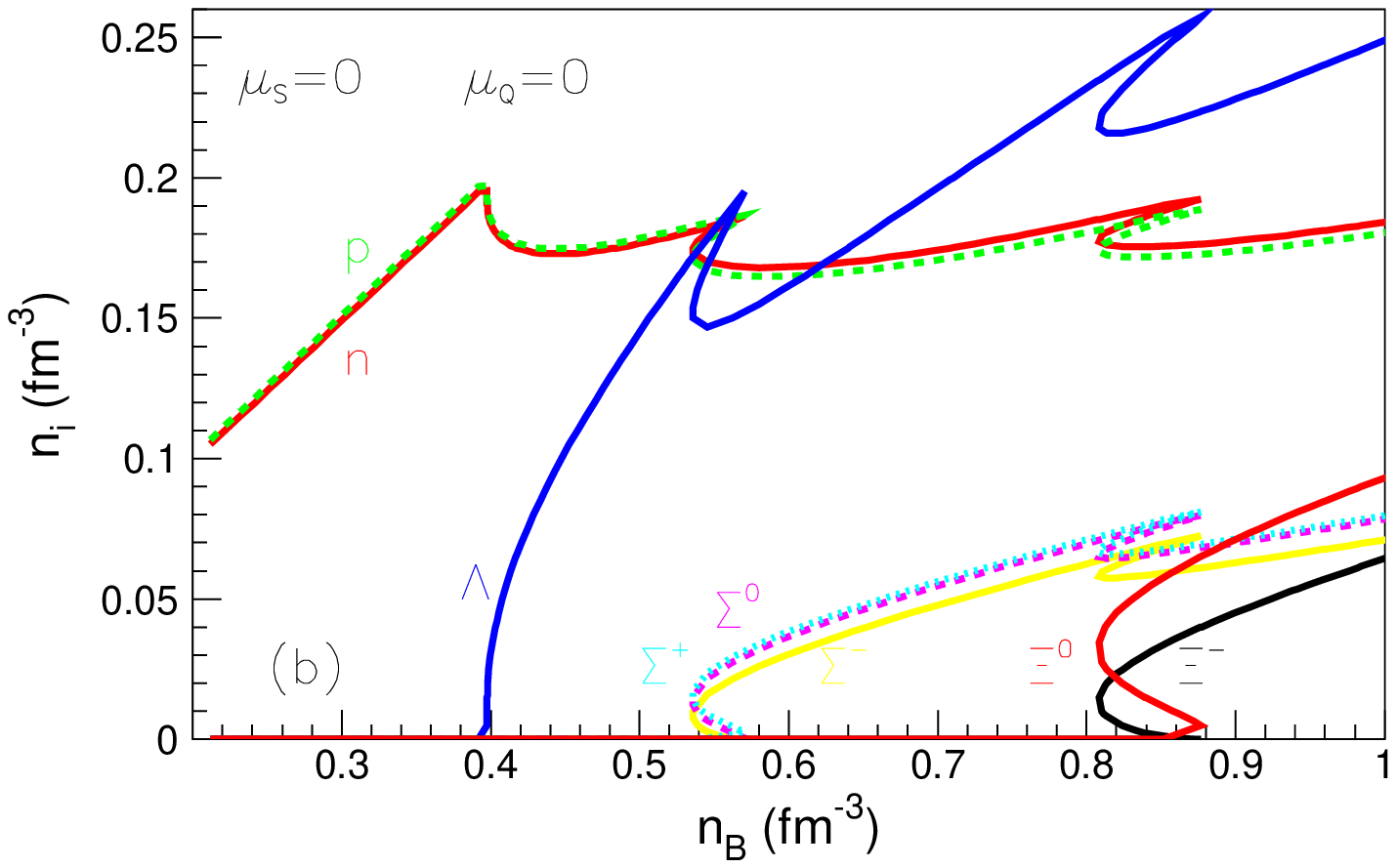}
\end{center}
\caption{(Color online)
Baryonic chemical potential (top) and particle abundances (bottom) 
as a function of baryon number density for $\mu_S=0$ and $\mu_Q=0$ at $T$=1 MeV,
employing the BGI parameterization~\cite{Balberg97}.
}
\label{fig:mu-rho}
\end{figure}

In order to highlight the effect of the Coulomb interaction, we will
start within this section by analyzing the phase diagram of pure
baryonic matter including the complete baryon octet, but without electrons.
The upper panel of Fig.~\ref{fig:mu-rho} illustrates the evolution of
the baryonic chemical potential as a function of baryon number density
at constant values of $\mu_S=0$, $\mu_Q=0$ and $T$=1
MeV~\footnote{This temperature value has been chosen for computational
  convenience.  The presented results are very close to the zero
  temperature limit.}. Three back-bending regions exist, which are all
correlated with a particle threshold, see the bottom panel of
Fig.~\ref{fig:mu-rho} where the corresponding particle abundances are
displayed. Since within the model of Balberg and Gal~\cite{Balberg97}
the hyperon-hyperon ($YY$)-interaction depends only weakly on the
particular channel, the order of the particle thresholds is mainly
given by their respective rest masses and chemical potentials. In the
present case, $\mu_S = \mu_Q = 0$, this means that the $\Lambda$ is
the first one to appear, followed by the almost degenerate
$\Sigma$-hyperons and then the Cascades. An investigation of $\bar
f(n_B)$ confirms that any back-bending can be cured by a Maxwell
construction and that the mixture of stable phases lowers the free
energy compared with the individual phases and corresponds thus to the
energetically favored solution. This means that three distinct phase
coexistence regions exist, induced by the onset of each hyperonic
family.

\begin{figure}
\begin{center}
\includegraphics[angle=0, width=0.99\columnwidth]{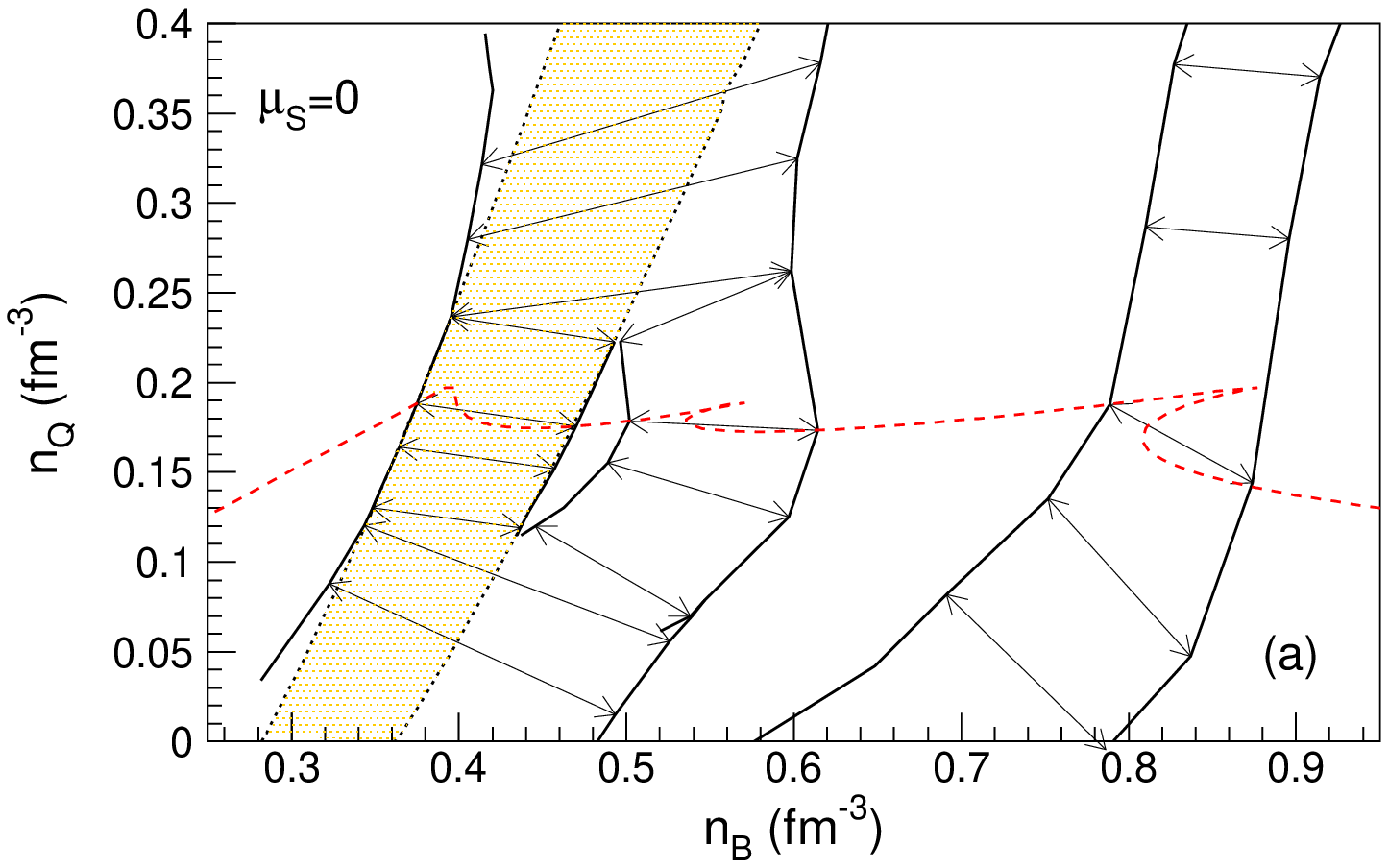}
\includegraphics[angle=0, width=0.99\columnwidth]{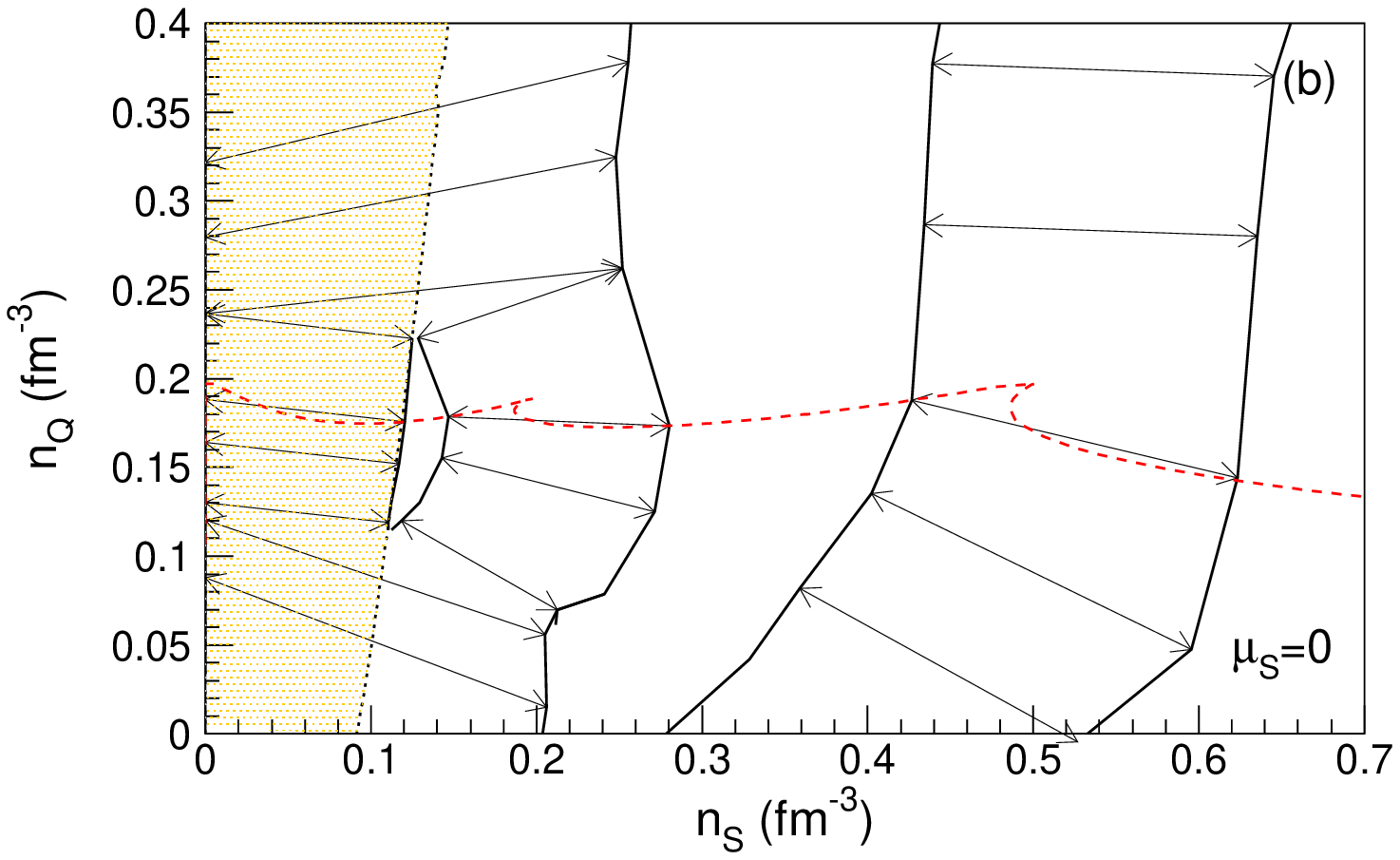}
\end{center}
\caption{(Color online)
Phase diagram of the 
$(n, p, Y)$-system 
for $\mu_S = 0$ 
at $T$=1 MeV as provided by the BGI parameterization~\cite{Balberg97}
in the $n_B$-$n_Q$ (a) and $n_S$-$n_Q$ (b) planes.
The arrows indicate the directions of phase separation. The $\mu_Q=0$ trajectory is illustrated with a (red) dashed line.
The hatched yellow area corresponds to the strangeness driven phase transition
domain of the simpler $(n,p,\Lambda)$-system studied in Ref. \cite{Gulminelli:2013qr}.  }
\label{fig:phd}
\end{figure}

Different thermodynamical conditions, i.e. different values of
$(\mu_S,\mu_Q)$ and $T$, will obviously change the particle
thresholds, their abundances, and the location of phase coexistence
regions.  By varying $\mu_S$ and $\mu_Q$, the whole 3-dimensional
phase diagram for a given temperature could be explored. Since we are
mostly interested here in astrophysical systems which can be
considered in strangeness changing weak equilibrium, $\mu_S=0$, we
will limit the analysis to this case.  The corresponding projections
of the phase diagram on the $n_B$-$n_Q$ (panel (a)) and the
$n_S$-$n_Q$-plane (panel (b)) are represented in Fig.~\ref{fig:phd}.
The arrows indicate the directions of phase separation.  In general,
increasing $\mu_Q$ leads to higher values of $n_Q$. The $\mu_Q=0$ trajectory considered in Fig.~\ref{fig:mu-rho} is here
represented with dashed lines. For the sake of completeness,
also the phase coexistence region of the $(n,p,\Lambda)$ systems
at supra-saturation densities is illustrated
as a hatched yellow area~\cite{Gulminelli:2012iq,Gulminelli:2013qr}.

Most of the time, the phase coexistence domains related to the onset
of $\Lambda$- and $\Sigma$-hyperons merge to a single domain extending
over a very large range of baryon number density, whereas the Cascade
thresholds remain separated leading to a second distinct phase
transition region.  This can be understood as follows: nonzero values
of $\mu_Q$, positive or negative, lift the degeneracy of the
$\Sigma$-hyperons, favoring the appearance of charged ones, negative
or positive depending on the sign of $\mu_Q$. The corresponding
threshold for $\Sigma^+$ or $\Sigma^-$ is shifted to lower densities
with increasing absolute value of $\mu_Q$, closer to the threshold of
neutral $\Lambda$-hyperons. At some critical value, the two phase
coexistence regions existing at $\mu_Q = 0$ merge into one single
domain.

As expected, the component of the order parameter in $n_Q$-direction
is more important if the number of charged particles participating in
the phase coexistence is large, i.e. for large values of $\mu_Q$. In
fact, although the total charge $n_Q$ is rather small for large
$\mu_Q$, the abundances of charged baryons itself are large. For
instance, the transition induced by $\Lambda$- and $\Sigma$-hyperons
shows an almost vanishing component in $n_Q$-direction close to $\mu_Q
= 0$, where three phase coexistence regions exist and the
$\Sigma$-onset is almost degenerate, see Fig.~\ref{fig:phd}, whereas
it becomes larger with increasing $|\mu_Q|$, i.e. in the regions where
only two phase coexistence domains exist. The $\Xi$-induced phase
transition has an order parameter with important contribution along
$n_Q$ whenever both $\Xi^0$ and $\Xi^-$ are created as their total
charge cannot vanish.  With increasing $\mu_Q$ the $\Xi^-$ production
threshold is shifted to higher densities, finally leaving the
considered density domain. Consequently, the charge dependence of the
order parameter becomes very weak for large $n_Q$.

\subsubsection{Coulomb effects on the phase diagram}
\label{section:neutral}

We now turn to investigate the influence of Coulomb effects on the
phase diagram. For simplicity, we will consider only electrons and
neglect other charged leptons or mesons.  The total free energy can be
written as the sum of a baryonic, leptonic and photonic contribution,
$f=f_B+f_L+f_{\gamma}$, where leptons and baryons are coupled only via
the strict electrical neutrality condition, $n_Q = n_L$. Leptons and
photons are well described by, respectively, fermionic and bosonic
ideal gases~\cite{Lattimer:1991nc}. 
None of them affects the phase structure.

\begin{figure}
\begin{center}
\includegraphics[angle=0, width=0.99\columnwidth]{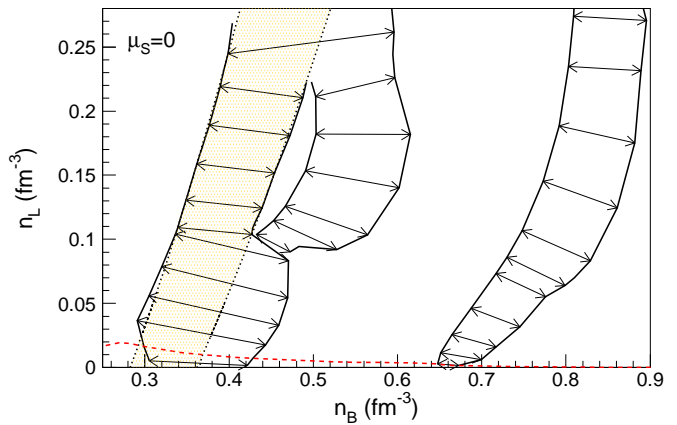}
\end{center}
\caption{(Color online)
Phase diagram of the $(n,p,Y, e)$-system at $T$=1 MeV for $\mu_S = 0$
as provided by BGI parameterization \cite{Balberg97} in the $n_B$-$n_L$-plane.
The (red) dashed curve marks the path corresponding to
$\beta$-equilibrium. The yellow hatched band marks the phase coexistence region obtained within the simpler $(n,p,\Lambda,e)$-system~\cite{Gulminelli:2013qr}. 
}
\label{fig:phd_+e}
\end{figure}

In the case of the $n,p,\Lambda$-system, where the order parameter has
only a small component in charge direction, the effect of electrons on
the phase diagram is small~\cite{Gulminelli:2013qr}. As discussed in
the previous section, upon including the full octet with all charged
hyperons, the dependence on the charged component becomes larger and
we thus expect more important Coulomb effects on the phase
diagram. This is confirmed by the results, see Fig.~\ref{fig:phd_+e},
where the phase diagram is displayed in the plane $n_B$-$n_L$. As
before, the arrows mark the directions of the order parameter. For the
sake of completeness the phase coexistence domain of the simpler
$(n,p,\Lambda,e)$-system is represented as a hatched yellow area.  The
qualitative structure of the phase diagram is similar to the case
without electrons discussed before, see Fig.~\ref{fig:phd}. It does
not come as a surprise, however, that the phase coexistence region
extents over a much smaller range in $n_B$, i.e. the phase transition
region is quenched by Coulomb effects, in particular, for low
$n_L$-values, where the the number of charged baryons participating in
the phase transition is large. Compared with the case without
electrons, the direction of phase separation is rotated in order to
reduce the difference in $n_L = n_Q$ between the two phases. The
reason is the large incompressibility of electrons, effectively
suppressing electron density fluctuations. For matter in
$\beta$-equilibrium, relevant for cold neutron stars, a large density
domain is actually covered by phase coexistence, roughly between $0.3
\lsim n_B \lsim 0.43$ fm$^{-3}$, see the dashed line in
Fig.~\ref{fig:phd_+e}.

As easy to anticipate, for small values of $\mu_Q$, corresponding to
the largest $n_L$-values shown in the figure, where the $\Lambda$ and
$\Sigma$ production thresholds are sufficiently different for the
corresponding phase coexistence regions to be separated, the
coexistence domain associated to the onset of $\Lambda$s in the
complete system sits exactly on the top of that corresponding to the
$(n,p,\Lambda,e)$-system. At smaller $\mu_L$ values, i.e. for small
values of $n_L$, the two systems have different phase coexistence
regions with the one of the full baryonic octet wider than that of the
simpler mixture due to the participation of $\Sigma$-hyperons.  In
particular, in the left bottom part of Fig.~\ref{fig:phd_+e} the low
density boundary of the phase coexistence region of the
$(n,p,Y,e)$-system is distinct from that corresponding to the
$(n,p,\Lambda,e)$-mixture.  There are two reasons for that.  For the
lowest considered $n_L$-values, the low density phase of the
restricted system is purely nucleonic while, upon considering the full
baryonic octet, the low density phase contains $\Sigma^-$, too, the
first hyperons to appear in that case.  At slightly higher but still
low $n_L$-values, the discrepancy arises from the composition of the
high density phase, which contains $\Sigma$-hyperons in addition to
nucleons and $\Lambda$-hyperons.

%%%%%%%%%%%%%%%%%%%%%%%%%%%%%%%%%%%%%%%%%%%%%%%%%%%%%%%%%%%%%%%%%%%%%%%%%
%%%%%%%%%%%%%%%%%%%%%%%%%%%%%%%%%%%%%%%%%%%%%%%%%%%%%%%%%%%%%%%%%%%%%%%%%%%%
It is important to stress that the phase diagram will
  obviously depend on the assumed coupling constants in all
  channels. For instance, if a repulsive potential is employed for
  $\Sigma$-hyperons, as suggested by the analyses of $(\pi^-,K^+)$-spectra,
  their threshold will be shifted to higher densities and
  the associated phase transition will consequently be shifted.  

\begin{figure}
\begin{center}
\includegraphics[angle=0, width=0.9\columnwidth]{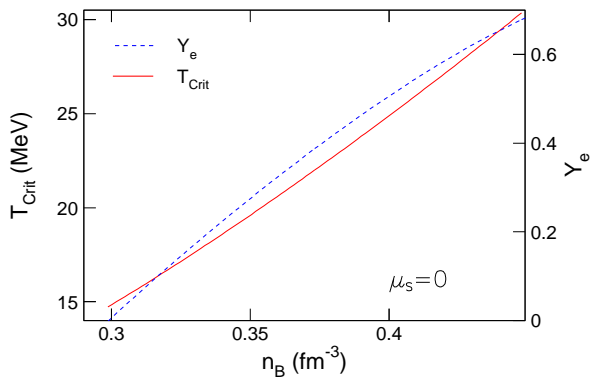}
\caption{(Color online) 
Electron fraction, $Y_e$, and $n_B$ at the corresponding critical
temperature for $\mu_S=0$ within the $(n,p,\Lambda,e)$-system.
Figure taken from Ref.~\cite{Gulminelli:2013qr}. 
\label{fig:critical}}
\end{center}
\end{figure}

For the simpler $n,p,\Lambda$+$e$-system, it has been shown that this
phase transition persists in this model at finite temperature, with a phase
separation direction almost independent of
$T$~\cite{Gulminelli:2013qr}. Increasing the temperature, the
  width of the coexistence region, shown for $T=0$ in
  Fig.~\ref{fig:phd_+e}, at low values of $n_L$ shrinks, leading to
  the appearence of a critical point above some finite value of $T$
of the order 15 MeV which survives up to very high
temperature. It moves to higher $n_L$-values with increasing
  temperature, see Fig.~\ref{fig:critical}, where the critical
temperature and the corresponding electron fraction $Y_e = n_L/n_B$
are shown. Since the character of the phase transition remains ``strangeness-driven'', 
we do not expect any qualitative change upon including the
full octet, with quantitative values of the same order.  These values
are typically reached within the cooling proto-neutron star, meaning
that effects of criticality should be experienced if there is a
strangeness-driven phase transition.  

In particular, the neutrino mean free path, $\lambda$, due to scattering off 
baryons should be strongly influenced. In order to explore this point,
we show calculations of the latter including the long-range correlations, essential for the
study of criticality. The linear response approximation is employed. 
In addition, since we focus here on the impact of density fluctuations close 
to the critical point, where spin-density fluctuations are expected to be 
small~\cite{Polls2002,Margueron2003}, only the vector channel will be considered. 
Then, in the non-relativistic limit for the baryonic components the mean free path
at temperature $T$ of a neutrino with initial energy $E_\nu$ is given
by \cite{Iwamoto1982,Navarro1999},
\begin{equation}
\frac{1}{\lambda} = \frac{1}{\lambda^{V}(E_\nu,T)} = \frac{G_F^2}{16\pi^2} \int (1+\cos\theta)
\mathcal{S}^{V}(q,T) (1-f_\nu(\mathbf{k}_3)) d\mathbf{k}_3.
\label{eq:lambdav}
\end{equation}
$G_F$ denotes here the Fermi constant, $\theta$ is the 
angle between the initial and final neutrino momentum (=$\mathbf{k_3}$),
$q$ is the transferred energy-momentum, $q=(\omega,\mathbf{q})$, and
$f_\nu$ is the Fermi-Dirac distribution of the outgoing neutrino.
$S^V$ represents the dynamical response function in the vector channel.
It is defined as 
\begin{eqnarray}
\mathcal{S}^{V}(q,T)&=&-\frac{2}{\pi}\frac{1}{1-\exp(-\omega/T)} \times
\nonumber \\
&&\left(\begin{array}{ccc} c_V^n & c_V^p & c_V^\Lambda \end{array} \right)
\Pi^{V}(q,T)
\left(\begin{array}{c} c_V^n\\ c_V^p \\ c_V^\Lambda \end{array} \right),
\end{eqnarray} 
where $\Pi^{V}(q,T)$ is the vector-polarization matrix for the three
species $n, p,$ and $\Lambda$. In mean field approximation it reduces
to the Lindhard functions~\cite{FetterWalecka} and in mean-field + RPA
approximation it is the solution of the Bethe-Salpeter
equations~\cite{Margueron2003,Iwamoto1982,Navarro1999}.  The coupling
constants, $c_V^i$, are $ -1 (n), 0.08 (p), -1
(\Lambda)$~\cite{Reddy1998}.  The residual p-h interaction entering
the calculation of the polarization matrix is closely related to the
curvature matrix without electrons \cite{Ducoin2008} and reflects
therefore the criticality of the phase transition.

\begin{figure}
\begin{center}
\includegraphics[angle=0, width=0.95\columnwidth]{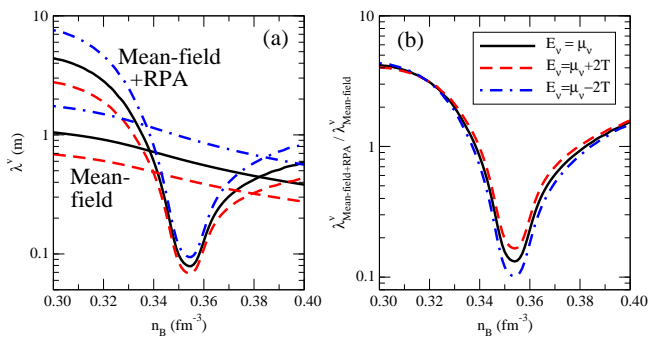}
\caption{(Color online) 
(a) Neutrino mean free path for the scattering off n, p, and $\Lambda$ at
$T=20$ MeV along a constant-$Y_e=0.2981$ trajectory in the phase diagram
for $E_\nu=\mu_\nu$, $\mu_\nu\pm T$ as a function of the
baryonic density, $n_B$.
The result of the mean-field approximation is compared with a mean-field+RPA calculation.
(b) The ratio of the mean free path within mean-field+RPA over mean-field
approximation is shown. Figure taken from Ref.~\cite{Gulminelli:2013qr}. 
\label{fig:mean_free_path}} 
\end{center}
\end{figure}

In Fig.~\ref{fig:mean_free_path} we show the calculation within the
$n,p,\Lambda,e$-system \footnote{Electrons are not relevant for this calculation.}. 
The depletion of the mean-free path in
the vicinity of the critical point is clearly visible. The possible
impact of a phase transition on the hydrodynamics of core collapse
will be discussed in Sec.~\ref{sec:astro}.

%%%%%%%%%%%%%%%%%%%%%%%%%%%%%%%%%%%%%%%%%%%%%%%%%%%%%%%%%%%%%%%%%%%%%%%%%%
%%%%%%%%%%%%%%%%%%%%%%%%%%%%%%%%%
%%%%%%%%%%%%%%%%%%%%%%%%%%%%%%%%%%%%%%%%%%%%%%%%

\section{Thermal effects on the EoS including non-nucleonic degrees of freedom}
\label{sec:sneos_non_uniform_additional}

In heavy-ion collisions (HICs), core-collapse supernovae and neutron
star or black-hole neutron star mergers, matter is strongly heated and
thermal effects on the EoS become important. Of course, the conditions
are very different in HICs compared with the astrophysical events. In
particular, matter in HICs cannot be considered in strangeness
changing weak equilibrium, but rather the zero net strangeness of the
two colliding nuclei is conserved during the entire collision due to
the very short timescales. However, thermal effects lead to a
considerable production of non-nucleonic degrees of freedom, hyperons,
nuclear resonances, strange and non-strange mesons, and in the early
phase probably a quark-gluon plasma. This clearly shows the importance
of thermal effects on the composition of matter. Therefore in compact
star astrophysics, where in addition to being heated, matter is
compressed to densities above nuclear matter saturation density,
$n_0$, non-nucleonic degrees of freedom are expected to occur, too.

This has been recognized shortly after the first discussions about
hyperons and quarks in cold neutron stars, and EoSs have been
developed to study the impact of additional particles, such as
hyperons and mesons, as well as meson condensate and quark matter
formation on the evolution of proto-neutron stars, see
e.g.~\cite{Prakash:1996xs} for an early review.  Inhomogeneous matter
in the outer layers and the formation of the neutron star crust do not
influence strongly the proto-neutron star evolution, such that in a
first approximation it can be neglected.  Therefore most of the works
investigating thermal effects on the EoS, treat homogeneous matter at
some given temperature or entropy per baryon and hadronic charge
fraction, $Y_Q = n_Q/n_B$, with values relevant for proto-neutron star
evolution, see e.g.~\cite{Pons:2000xf,Pons:2000iy,Bombaci:2011mx,Menezes:2007hp,Dexheimer:2008ax,yasutake09}.  Only
recently, some models have been proposed, including non-nucleonic
degrees of freedom and treating the low-density and low-temperature
inhomogeneous nuclear matter, too, see e.g.~\cite{Oertel:2012qd,sagert_09,Ishizuka_08,nakazato08}.

Quark matter formation is very interesting in this context, in
particular if the transition occurs during the early post-bounce phase,
since then the related phase transition is visible in the supernova
neutrino signal~\cite{sagert_09}. Such an early phase transition
requires, however, a low transition density and therefore a relatively
soft EoS which seems difficult to be reconciled with a maximum cold
neutron star mass above $\sim 2 M_\odot$ in agreement with recent
observations~\cite{Antoniadis_13,Demorest_10}, see
Ref.~\cite{sagert_12}. In the contracting and partially deleptonized
proto-neutron star, higher densities and temperatures are reached, and
the conditions are, therefore, more favorable for quark matter to
appear. Such a transition could explain some gamma-ray-bursts or, via
the scenario of so-called ``quark-novae'', some unusual supernova
light-curves, see the discussion in Ref.~\cite{Buballa_14} and
references therein.  In addition, quark matter at not too high
temperatures has a very rich phase structure related to different
color superconducting phases with consequences for compact star
phenomenology, for instance on neutron star cooling, see e.g.~\cite{Alford07}
for a review.

Here, we will concentrate on hadronic degrees of freedom,
i.e. hyperons and mesons, and present some selected EoS in order to
illustrate thermal effects on the appearance of these particles. Due
to the computational complexity, only some microscopic calculations
exist, see e.g. the BHF calculations
of~\cite{nicotra_06,Burgio11,chen_12}. Most models employ
phenomenological interactions, non-relativistic or relativistic mean
field (RMF) ones. In the following section we will, therefore, briefly
introduce the latter, the former has already been discussed in
Sec.~\ref{section:uncharge}. 

Pions, and to less extent kaons, have already been considered twenty
years ago as possible candidates for the hot and dense matter in
supernova cores and proto-neutron stars, see
e.g.~\cite{Pons:2000xf,Pons:2000iy,Mayle93}. The authors of
Ref.~\cite{Mayle93} argue that the temperature of the supernova core
could be increased by the presence of pions. As a consequence, the
number of electron neutrinos would increase resulting in a higher
neutrino luminosity favoring a successful explosion. However, the
employed pion-nucleon interaction is probably too attractive, and a
more realistic interaction decreases the number of pions eventually
present in supernova cores and thus the effect on the neutrino
luminosity. For cold neutron stars, it is now commonly assumed that
there is an $s$-wave $\pi N$ repulsion, preventing pions from forming
a Bose-Einstein condensate. Most recent works on the EoS for hot and
dense matter neglect for simplicity any interaction and use a free
pion gas~\cite{Oertel:2012qd,Peres_13,Ishizuka_08,nakazato08}. This
will be the case for the results shown below, too. At least for the
high temperatures and low densities thermal effects should dominate
and the interaction should be less important. Let us mention that in
Ref.~\cite{Ishizuka_08} it is shown that a parameterized $s$-wave $\pi
N$ repulsive interaction allows to avoid pion condensation at low
temperatures but, since the effects of the interaction on the EoS are
only important at low temperatures, and pionic effects are generally
not the dominant contribution to the EoS, the model dependence of the
results becomes very weak.

\subsection{Relativistic mean field models for the EoS}
The literature on phenomenological RMF models is large and many
different versions exist (see e.g.~\cite{Dutra:2014qga}). The basic idea
is that the interaction between baryons is mediated by meson
fields. These are not real mesons, but introduced on a
phenomenological basis with their quantum numbers in different
interaction channels. Earlier models introduce non-linear
self-couplings of the meson fields in order to reproduce correctly
nuclear matter saturation and properties of nuclei, whereas more
recently density-dependent couplings between baryons and the meson
fields have been widely used.  The Lagrangian of the model can be
written in the following form
%%%%%%%%%%%%%%%%%%%%%%%%%%%%%%%%%
\begin{eqnarray}
{\mathcal L} &=& \sum_{j \in \mathcal{B}}  \bar \psi_j \left( i \gamma_\mu \partial^\mu - m_j +
  g_{\sigma j} \sigma
  + g_{\sigma^* j} \sigma^* \right.\nonumber \\ &&\left. + g_{\delta
    j} \vec{\delta} \cdot \vec{I}_j - g_{\omega j} \gamma_\mu \omega^\mu - g_{\phi j}
\gamma_\mu \phi^\mu - g_{\rho j} \gamma_\mu \vec{\rho}^\mu \cdot \vec{I}_j\right) \psi_j \nonumber \\ &&
 + \frac{1}{2} (\partial_\mu \sigma \partial^\mu \sigma - m_\sigma^2 \sigma^2)
  - \frac{1}{3} g_2 \sigma^3 - \frac{1}{4} g_3 \sigma^4 \nonumber \\ && 
 + \frac{1}{2} (\partial_\mu \sigma^* \partial^\mu \sigma^* - m_{\sigma^*}^2
  {\sigma^*}^2) \nonumber \\ &&
 + \frac{1}{2} (\partial_\mu \vec{\delta} \partial^\mu \vec{\delta} - m_{\delta}^2
  {\vec{\delta}}^2) \nonumber \\ &&
- \frac{1}{4}
W^\dagger_{\mu\nu} W^{\mu\nu} 
- \frac{1}{4}
P^\dagger_{\mu\nu} P^{\mu\nu} 
- \frac{1}{4}
\vec{R}^\dagger_{\mu\nu} \cdot \vec{R}^{\mu\nu} \nonumber \\ && 
+ \frac{1}{2} m^2_\omega \omega_\mu \omega^\mu + \frac{1}{4} c_3 (\omega_\mu \omega^\mu)^2 \nonumber \\ && 
+ \frac{1}{2} m^2_\phi \phi_\mu \phi^\mu 
+ \frac{1}{2} m^2_\rho \vec{\rho}_\mu \cdot \vec{\rho}^\mu ~,
\end{eqnarray}
where $\psi_j$ denotes the field of baryon $j$, and $W_{\mu\nu}, P_{\mu\nu}, \vec{R}_{\mu\nu}$ are the
vector meson field tensors of the form
\begin{eqnarray}
V^{\mu\nu} = \partial^\mu  V^\nu - \partial^\nu V^\mu~.
\end{eqnarray}
$\sigma, \sigma^*$ are scalar-isoscalar  meson
fields, coupling to all baryons ($\sigma$) and to strange baryons
($\sigma^*$), respectively. $\vec{\delta}$ induces a scalar-isovector
coupling. 

In mean field approximation, the meson fields are
replaced by their respective mean-field expectation values, which are given in
uniform matter as
\begin{eqnarray}
m_\sigma^2 \bar\sigma + g_2 \bar\sigma^2 + g_3 \bar\sigma^3 &=& \sum_{i \in B} g_{\sigma i} n_i^s
\\
m_{\sigma^*}^2 \bar\sigma^* &=& \sum_{i \in B} g_{\sigma^* i} n_i^s\\
m_\delta^2 \bar\delta &=& \sum_{i \in B} g_{\delta i} t_{3 i} n_i^s\\
m_\omega^2 \bar\omega + c_3 \bar\omega^3 &=& \sum_{i \in B} g_{\omega i} n_i\\
m_\phi^2 \bar\phi &=& \sum_{i \in B} g_{\phi i} n_i\\
m_\rho^2 \bar\rho &=& \sum_{i \in B} g_{\rho i} t_{3 i} n_i~,
\end{eqnarray}
where  $\bar\delta=\langle\delta_3\rangle$, $\bar\rho=\langle\rho_3^0\rangle$,
$\bar\omega=\langle\omega^0\rangle$, $\bar \phi=\langle\phi^0\rangle$, and
$t_{3 i}$ represents the third component of isospin of baryon $i$ with the convention that
$t_{3 p} = 1/2$. The scalar density of baryon $i$ is given by 
\begin{equation}
n^s_i = \langle \bar \psi_i \psi_i \rangle = \frac{1}{\pi^2} \int
k^2 \frac{M^*_i} {\sqrt{k^2 + M^{*2}}} \{f[\epsilon_i(k)] +\bar{f}[\epsilon_i(k)]\} dk~,
\end{equation}
and the number density by 
\begin{equation}
n_i = \langle \bar \psi_i\gamma^0 \psi_i \rangle = \frac{1}{\pi^2} \int
k^2 (f(\epsilon_i(k)) - \bar{f}(\epsilon_i(k)))dk ~.
\end{equation}
$f$ and $\bar{f}$ represent here the occupation numbers of the
respective particle and antiparticle states with the single-particle
energies, $\epsilon_i(k) = \sqrt{k^2 + M^{*2}}$, which reduce to a step
function at zero temperature. The effective baryon
mass $M^*_i$ depends on the scalar mean fields as
\begin{equation}
M^*_i = M_i - g_{\sigma i} \bar\sigma - g_{\sigma^* i} \bar\sigma^* -
g_{\delta i} t_{3 i} \bar \delta~,
\end{equation}
and the effective chemical potentials, $(\mu_i^*)^2 = (M_i^*)^2 + k_{Fi}^2$,
are related to the chemical potentials via
\begin{equation}
\mu_i^* = \mu_i - g_{\omega i} \bar\omega - g_{\rho i} \,t_{3 i}
\bar\rho - g_{\phi i} \bar \phi - \Sigma_0^R~.
\label{mui}
\end{equation} 
The rearrangement term 
\begin{eqnarray}
\Sigma_0^R &=& \sum_{j \in B} \left( \frac{\partial g_{\omega j}}{\partial n_j}
\bar\omega n_j + t_{3 j} \frac{\partial g_{\rho j}}{\partial n_j}
\bar\rho n_j +\frac{\partial g_{\phi j}}{\partial n_j}
\bar\phi n_j \right. \nonumber \\ && \left. -\frac{\partial g_{\sigma j}}{\partial n_j}
\bar\sigma n_j^s -\frac{\partial g_{\sigma^* j}}{\partial n_j}
\bar\sigma^* n_j^s - t_{3 j} \frac{\partial g_{\delta j}}{\partial n_j}
\bar\delta n_j^s \right)~.
\end{eqnarray}
is present in density-dependent models to ensure thermodynamic consistency. 

In the present paper we consider a set of models  frequently used
  in the literature that succeed in
describing  a 2$M_\odot$ neutron star. In particular, we will show
results for three non-linear
models, GM1~\cite{glendenning_91}, TM1~\cite{Sugahara_94} and
TM1-2~\cite{providencia13}, and two density-dependent ones,
DDH$\delta$~\cite{Avancini2009,Gaitanos} and DD2~\cite{Typel:2009sy}. 
For
the GM1 parameterization, $c_3$ =0, and the $\delta$-field is absent in
GM1, TM1, TM1-2 and DD2.  The TM1-2 parametrization has
  properties similar to TM1 at saturation but is stiffer at large
  densities. TM1-2 has a quite large symmetry energy slope $L$ at
  saturation ($L=110$ MeV), therefore we have consired a modification
  with a smaller slope, $L=55$ MeV,  by including a
  $\omega\rho$ mixing term which allows us to discuss the
  influence of the density dependence of the symmetry energy on the
  results. In Table \ref{tab:nuclear} we show two parametrizations for
  TM1-2 corresponding to the two values of $L$.
The density-dependent models assume $g_2 = g_3 = c_3 =
0$ (no non-linear terms) and the couplings become density dependent, 
\begin{equation}
g_i(n_B) = g_i(n_0) h_i(x)~,\quad x = n_B/n_0~.
\end{equation}
There exist different parameterization employing mostly the same
functional forms.  Within the DDH$\delta$ and the DD2
parameterization, the following forms are assumed for the isoscalar
couplings,
\begin{equation}
h_i(x) = a_i \frac{1 + b_i ( x + d_i)^2}{1 + c_i (x + d_i)^2}
\end{equation}
and 
\begin{equation}
h_i(x) = a_i\,\exp[-b_i (x-1)] - c_i (x-d_i)~.
\end{equation}
for the isovector ones.  In the following we will show results
  for extensions of the STOS~\cite{shen_98} and the statistical model
  EoS by Hempel and Schaffner-Bielich~\cite{Hempel09} (BHB model),
  that have as underlying parametrizations, respectively, TM1 and DD2,
  and for the non-relativistic EOS, LS220 \cite{LS}.
  As extra degrees
  of freedom pions, $\Lambda$-hyperons, or all hyperons of the
  baryonic octet have been considered, and accordingly, we will add
  the termination $\pi$, $\Lambda$ or $Y$ to the name othe EoS. The
  original references are given in Table~\ref{tab:hypcouplings},
  together with the hyperonic interaction employed. 

The wealth of nuclear data allows to constrain reasonably the
parameter values of the interaction between nucleons. The
corresponding parameter values of the different models can be found in
the above references and the resulting nuclear matter properties are
listed in Table~\ref{tab:nuclear}. In addition to standard properties
of isospin symmetric nuclear matter, the energy per baryon of pure
neutron matter at saturation density is given, too, for which recently
a range
\begin{equation}
14.1 \lsim E/A(n_0)\lsim 21.4 \mathrm{MeV}.
\label{Kruger}
\end{equation}
has been derived from microscopic calculations within chiral nuclear
forces~\cite{Kruger_13}. This quantity is particularly interesting for
the EoS of compact stars, completing the information about symmetric
matter, since very asymmetric matter close to pure neutron matter is
encountered. Except for the DDH$\delta$-model, for which the value is
too low, the employed models lie within the indicated range. The
symmetry energy $E_{sym}$ and its slope, $L$, containing information
about the isospin dependence of the EoS, too, are as well important in
this respect. For instance, it is well known that the radius of
compact stars is very sensitive to
$L$~\cite{Horowitz:2001ya,Carriere:2002bx,rafael11}. The models considered here span a
wide range of values of $L$, with some of them being at the upper end
of possible values~\cite{Lattimer_13}.

%%%%%%%%%%%%%%%%%%%%%%%%%%%%%%%%%%%%%%%%%%%%%%%%%%%%%%%%%%%%%%%%%%%%%%%%%%%%%
\begin{table*}[t!]
\begin{center}
 \begin{tabular}{c|cccccc}
  &$K$& $E_\mathit{sym}$ & $n_0 $& $B$ 
  &$L$  &$E/A(n_0)$\\ 
& [MeV] & [MeV] & [MeV] & [MeV]& [ MeV] & [ MeV] \\
\hline \hline
GM1 &300&32.5&0.153&16.3&94 &18.6\\
TM1&281 &36.9 &0.145 &16.3 &111 & 21.1\\
TM1-2&282&37.2&0.146&16.4&111/55& 21.8/17.1 \\
DD2 &243&31.7 &0.149 &16.0&55 & 18.2 \\
DDH$\delta$ &240&25.1 &0.153&16.3&44 & 10.6\\
LS220 &220&28.6 &0.155&16.0&74 &14.4 \\
\end{tabular}
\caption{Nuclear matter properties of the models considered in
  this study for symmetric nuclear matter at saturation, except for the last
  column where  the energy per baryon of neutron matter at $n_0$
is given with the neutron mass subtracted.}
\label{tab:nuclear}
\end{center}
\end{table*}
%%%%%%%%%%%%%%%%%%%%%%%%%%%%%%%%%%%%%%%%%%%%%%%%%%%%%%%%%%%%%%%%%%%%%%%%%%%%%%%
%%%%%%%%%%%%%%%%%%%%%%%%%%%%%%%%%%%%%%%%%%%%%%%%%%%%%%%%%%%%%%%%
\begin{table*}
\begin{tabular}{l|ccccccccc|l}
\hline
 Model& Nuclear &$R_{\sigma^* \Lambda}$
&$ R_{\sigma^* \Xi}$ &$R_{\sigma^* \Sigma}$& 
$R_{\omega Y}$ 
&$R_{\phi Y}$ 
&$U_{\Lambda}^{(\Lambda)}(n_0/5) $  &
$U_{\Xi}^{(\Xi)}(n_0/5) $ &
$U_{\Sigma}^{(\Sigma)}(n_0/5) $ & Reference(s)
\\
&interaction& &&&&& [MeV]&[MeV]&[MeV]& \\ \hline \hline
BHB$\Lambda$  &DD2&  0 &- &- & 1&0 & -5 &-  & - &\cite{Banik2014}\\ 
BHB$\Lambda\Phi$  &DD2& 0 &- &- & 1&1 &7 &- &- &\cite{Banik2014}\\ 
STOS$\Lambda$  & TM1&0 &-  &-  & 1&0 &6 & -  & - &\cite{Shen:2011qu}\\ 
LS220$\Lambda$ 
& LS220& - &- & - &- &- & -5&- &-&\cite{Gulminelli:2013qr,Peres_13}\\ 
\hline
STOSY  &TM1&0.67 &1.23 &0.67& 1&1 & -11& -8&5 &\cite{Ishizuka_08}\\ 
GM1 $Y6$  & GM1& 0 &0.55 &0 & 2&2 & -7&-10 & 13&\cite{Oertel_14}\\ \
TM1-2 Y1&TM1-2(111)& 0&0&0& 1& 1& 1.7&21.1 &16.2&\cite{Oertel_14}\\%& 0.00137 \\
TM1-2 Y2&TM1-2(55)& 0&0&0&  1& 1& 1.7&21.1 & 16.2&\cite{Oertel_14}\\ 
TM1-2 $\Lambda$4& TM1-2(111)&  1.68&1.68& 1.68& 1.5 & 2&-41.1&36.0 &-21&\cite{Oertel_14}\\ 
TM1-2 $\Lambda$6&TM1-2(55)& 1.58&1.58& 1.58&1.5 & 2&-33.7&44.4 &-12.8&\cite{Oertel_14}\\ 
DDH$\delta$ Y4  &DDH$\delta$ & 1.03 &0 &0  &1.5 &0.85 & -5&79 &62&\cite{Oertel_14}\\ 
220g3 
 & LS220 & - &  - &  - &  - &  - & -2.73 & -2.73 & -2.73 &\cite{Oertel:2012qd}\\

\hline

\end{tabular}
\caption{ Summary of coupling parameters used within the different
  models. The nuclear interaction is indicated in the second column
  and those of hyperons in columns 3-7. The couplings to the isoscalar
  vector mesons are defined with respect to the respective $SU(6)$
  values and the couplings to $\sigma^*$ are defined with respect to
  $g_{\sigma N}$. The coupling parameters of the two non-relativistic models, LS220$\Lambda$ and 220g3 are given in Table~\ref{tab:bg1}. In addition, the values of the hyperon single
  particle potentials in hyperon matter at $n_0/5$ are given for
  information.
The last column indicates the reference(s) to
  the original work. The models in the upper part thereby consider
  only $\Lambda$-hyperons whereas those in th lower part allow all
  hyperons to have nonzero abundances.  }
\label{tab:hypcouplings}
\end{table*}

%%%%%%%%%%%%%%%%%%%%%%%%%%%%%%%%%%%%%%%%%%%%%%%%%%%%%%%%%%%%%%%%

Hyperonic data are scarce and it is therefore very difficult to obtain
information on the interaction parameters in the hyperonic sector.
Many recent works, see
e.g.~\cite{Weissenborn11c,Banik2014,Miyatsu2013}, use a procedure
inspired by the symmetries of the baryon octet to reduce the number of
free parameters. The individual isoscalar vector meson-baryon
couplings can then be expressed in terms of $g_{\omega N}$ and a few
additional parameters, $\alpha, \theta, z = g_1/g_8$, see
e.g.~\cite{Schaffner:1995th} for details, as follows 
{\small
\begin{eqnarray}
\frac{g_{\omega\Lambda}}{g_{\omega N}} &=& \frac{ 1 - \frac{2 z}{\sqrt{3}} 
  (1-\alpha) \tan \theta}{1- \frac{z}{\sqrt{3}}  (1 - 4 \alpha)
  \tan\theta} ~,\  
\frac{g_{\phi\Lambda}}{g_{\omega N}} = -\frac{ \tan\theta + \frac{2 z}{\sqrt{3}} 
  (1-\alpha)}{1- \frac{z}{\sqrt{3}}  (1 - 4 \alpha)
  \tan\theta} ~,\nonumber \\
\frac{g_{\omega\Xi}}{g_{\omega N}} &=& \frac{ 1 - \frac{z}{\sqrt{3}} 
  (1+ 2 \alpha) \tan \theta}{1- \frac{z}{\sqrt{3}}  (1 - 4 \alpha)
  \tan\theta} ~,\ 
\frac{g_{\phi\Xi}}{g_{\omega N}} = -\frac{ \tan\theta + \frac{z}{\sqrt{3}} 
  (1+ 2 \alpha)}{1- \frac{z}{\sqrt{3}}  (1 - 4 \alpha)
  \tan\theta} ~,\nonumber \\
\frac{g_{\omega\Sigma}}{g_{\omega N}} &=& \frac{ 1 + \frac{2 z}{\sqrt{3}} 
  (1- \alpha) \tan \theta}{1- \frac{z}{\sqrt{3}}  (1 - 4 \alpha)
  \tan\theta} ~,\ 
\frac{g_{\phi\Sigma}}{g_{\omega N}} = \frac{ -\tan\theta + \frac{ 2 z}{\sqrt{3}} 
  (1- \alpha)}{1- \frac{z}{\sqrt{3}}  (1 - 4 \alpha)
  \tan\theta} ~,\nonumber \\
\frac{g_{\phi N}}{g_{\omega N}} &=& -\frac{ \tan\theta + \frac{z}{\sqrt{3}} 
  (1- 4 \alpha)}{1- \frac{z}{\sqrt{3}} (1 - 4 \alpha)
  \tan\theta} ~.
\label{eq:symmetry}
\end{eqnarray} }
The following values are commonly assumed: $\tan\theta = 1/\sqrt{2}$,
corresponding to ideal $\omega$-$\phi$-mixing, $\alpha = 1$, and $z =
1/\sqrt{6}$. The latter value reflects an underlying $SU(6)$-symmetry,
and only recent studies in view of the observation of high mass
neutron stars have relaxed this assumption, for
example~\cite{Weissenborn11c,Lopes2014,Ishizuka_08,Miyatsu2013}, or
even varied freely the hyperonic isoscalar vector couplings, see
e.g.~\cite{Oertel_14}.

In the isovector sector, not the same procedure is applied, since this would
lead to contradictions with the observed nuclear symmetry energy. $g_{\rho N}$
is therefore left as a free parameter, adjusted to the desired value of the
symmetry energy, and the remaining isovector
vector couplings are fixed by isospin symmetry. 

For the scalar sector, different methods are applied. In
Ref.~\cite{Colucci2013} a symmetry inspired procedure is discussed
together with the constraints imposed by hypernuclear data. In
\cite{Shen:2011qu}, the value of the $\sigma\Lambda$-coupling is taken
from a fit of the binding energies of single $\Lambda$-hypernuclei
resulting in $R_{\sigma\Lambda} = 0.621$. In many other works, see
e.g. Refs.~\cite{Weissenborn11c,Banik2014,Oertel_14,Ishizuka_08}, the
information from hypernuclear data on hyperonic single-particle mean
field potentials is used to constrain the coupling constants. Let us
emphasize that almost no information is available on the
hyperon-hyperon ($YY$)-interaction apart from a few light
double-$\Lambda$-hyper\-nuclei, that constrain only the low density
behavior.  Therefore, the corresponding couplings, within the RMF
models, in particular $\sigma^*$ and $\phi$ are very poorly
constrained. In most recent models $\phi$-mesons are added in order to
be compatible with the 2 $M_\odot$ neutron star, whereas often
$\sigma^*$ is neglected (see
e.g.~\cite{Bednarek:2011gd,Weissenborn11c,Banik2014}). This leads,
however, to a very repulsive $YY$-interaction already at very low
densities~\cite{Oertel_14,Ishizuka_08,Torres15} which does not appear
very realistic in view of the double-$\Lambda$-hypernuclear data.  In
Table~\ref{tab:hypcouplings} we summarize the coupling parameters used
within the different models employed here. We list, in addition, the
values of the corresponding hyperonic single particle potentials in
hyperonic matter.  Data on the bond energy of
double-$\Lambda$-hypernuclei can be reinterpreted in terms of the
$\Lambda$ potential in $\Lambda$ matter at the average density of
$\Lambda$ inside those nuclei~\cite{Vidana01,Khan15}. Mean-field
calculations have shown that in light nuclei (from He to C) the
average $\Lambda$ density is close to one fifth of the saturation
density~\cite{Vidana01,Khan15}. We have therefore chosen this density
for the reference values of the hyperonic potentials.

The model LS220$\Lambda$ presents a strangeness-driven phase
transition at the onset of
$\Lambda$-hyperons~\cite{Gulminelli:2012iq,Gulminelli:2013qr}, see
Sec.~\ref{section:PT}. Among the relativistic models, the parameter
sets TM1-2$\Lambda$4/TM1-2$\Lambda$6 are chosen at the limit of
presenting a thermodynamic instability at the onset of
$\Lambda$-hyperons and stable with respect to all other
hyperons~\cite{Oertel_14}. The other models do not show any
instability related to the onset of hyperons. To summarize, we believe
that the large set of employed models and coupling constants gives a
representative estimate of the present theoretical uncertainties in
the high density equation of state modelling at finite temperature.

%%%%%%%%%%%%%%%%%%%%%%%%%%%%%%%%%%%%%%%%%%%%%%%%%%%%%%%%%%%%%%%%
\begin{table}
\begin{center}
\begin{tabular}{l|cccc}
\hline
 Model& $M_{\mathit{max}}$ & $R_{1.4}$ & $f_S$& $n_B^{(c)}$  \\
&$[M_\odot]$ & [km]& & [fm$^{-3}$] \\ \hline \hline
GM1  & 2.39  &13.7  & 0 & 0.84\\ 
DD2         &2.42   & 13.2 & 0 &0.84 \\ 
TM1 (STOS)  & 2.23  &14.5  &0  &0.82   \\ 
TM1-2(L=111)  & 2.31   &14.3  & 0 &0.81  \\ 
TM1-2 (L=55)  & 2.24   &13.3  & 0& 0.86 \\ 
DDH$\delta$  & 2.16  & 12.6 &0 &0.98  \\ 
LS220  & 2.06  & 12.7 & 0&1.11   \\ 
\hline
GM1 $Y6$  & 2.29   &13.8  &0.04  &0.85  \\ 
BHB$\Lambda$  & 1.96  &13.2  &0.05 & 0.95  \\ 
BHB$\Lambda\Phi$  &2.11 &13.2&0.05& 0.95\\
STOS$\Lambda$  &1.91 &14.4&0.04 &0.88\\
STOSY  &1.65 &14.4 &0.07& 0.67 \\
STOSY$\pi$  &1.66 &13.6 &0.05 &0.81  \\
TM1-2  Y1 & 1.95 &14.6&0.15&0.86\\
TM1-2  Y2 & 1.94 &13.4&0.12&0.91\\
TM1-2 $\Lambda$4& 2.13&14.6 &0.16&0.90 \\ 
TM1-2 $\Lambda$6& 2.09&13.4 &0.11&0.92 \\ 
DDH$\delta$ Y4  & 2.05 &12.7&0.04&0.99 \\ 
LS220$\Lambda$  & 1.91 &12.4 &0.06&1.37 \\ 
LS220$\pi$  & 1.95 &12.2 & - & 1.27\\ 
220g3  & 1.95 &12.7 & 0.005 & 0.98\\ 
\hline

\end{tabular}
\caption{ Results calculated within different models at zero
  temperature: maximum mass of a cold spherical symmetric neutron star
  in $\beta$-equilibrium, radius at a fiducial mass of $M = 1.4
  M_\odot$, the total strangeness fraction, $f_S$, representing the
  integral of the strangeness fraction $Y_s/3$ over the whole star as
  in Ref.~\cite{Weissenborn11c}, and the central baryon number
  density. In the upper part purely nucleonic models are listed and in
  the lower part those containing hyperons and/or pions are given. }
\label{tab:nsresultsT0}
\end{center}
\end{table}

\subsection{Hyperons and pions in the finite temperature EoS}
%%%%%%%%%%%%%%%%%%%%%%%%%%%%%%%%%%%%%%%%%%%%%%%%%%%%%%%%
\begin{figure*}
\begin{center}
\includegraphics[width=0.7\textwidth]{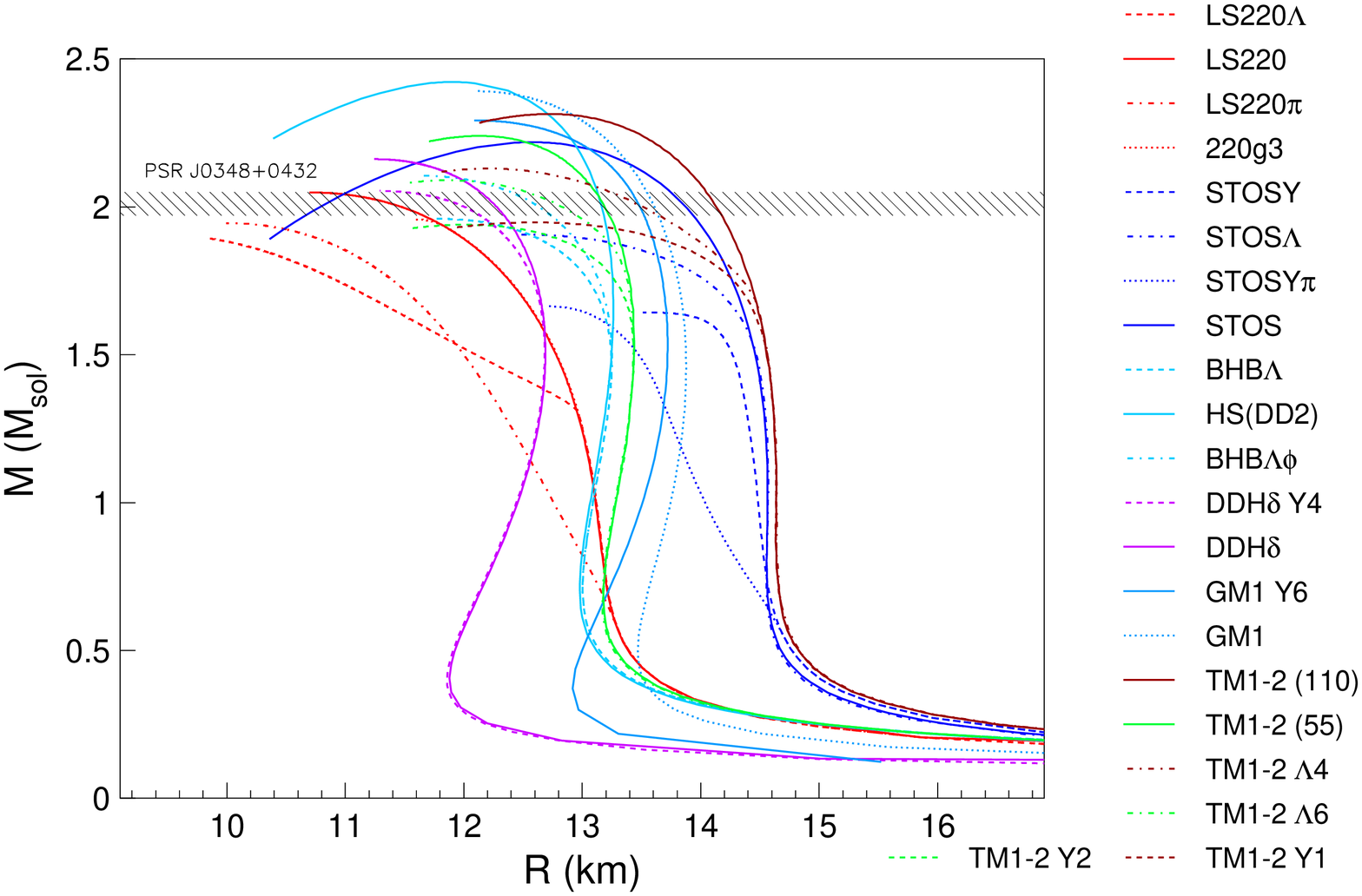} 
\caption{(Color online) Mass-radius relation of a spherically
  symmetric cold $\beta$-equilibrated neutron star for different EoS
  including hyperons and/or pions and the corresponding purely nuclear EoS. 
\label{fig:eos3d}}
\end{center}
\end{figure*}
%%%%%%%%%%%%%%%%%%%%%%%%%%%%%%%%%%%%%%%%%%%%%%%%%%%%%%%%%%%%%%
%%%%%%%%%%%%%%%%%%%%%%%%%%%%%%%%%%%%%%%%%%%%%%%%%%%%%%%%
\begin{figure}
\includegraphics[width=0.95\columnwidth]{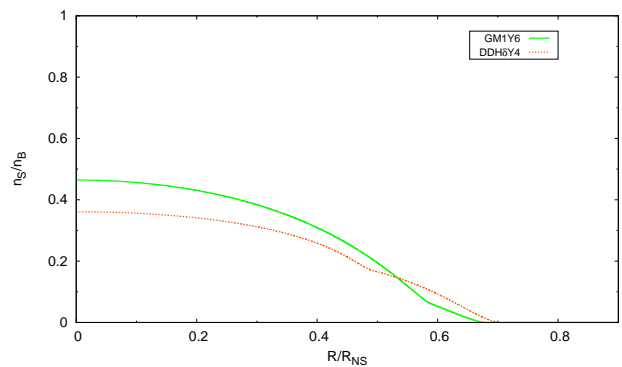} 
\caption{(Color online) 
  Strangeness fraction profiles of the maximum mass configurations
  within some of the EoS models shown in Fig.~\ref{fig:eos3d}.
\label{fig:hypfractions}}
\end{figure}
%%%%%%%%%%%%%%%%%%%%%%%%%%%%%%%%%%%%%%%%%%%%%%%%%%%%%%%%%%%%%%
%%%%%%%%%%%%%%%%%%%%%%%%%%%%%%%%%%%%%%%%
Let us start with some comments on the EoS of cold
$\beta$-equilibrated neutron stars including non-nucleonic degrees of
freedom, for more details see~\cite{Chatterjee_this_issue}. By simple
arguments based on the Pauli principle, for a system composed of
fermionic particles, additional degrees of freedom tend to soften the
EoS. In turn, this reduces the neutron star maximum mass, eventually
being in contradiction with the recently observed
masses~\cite{Antoniadis_13,Demorest_10}. The way out is of course that
the interaction must be much more repulsive at high density than
presently assumed.  This is true for hyperons and quarks. On the quark
side, this leads to the problem of reconfinement, i.e. the hadronic
EoS becomes again energetically favored at some very high
density~\cite{Zdunik:2012dj}, and on the hyperonic side the additional
repulsion leads in general to a very low strangeness content of
neutron stars~\cite{Weissenborn11c}, see \cite{Oertel_14} for models
with higher strangeness content.
 
%%%%%%%%%%%%%%%%%%%%%%%%%%%%%%%%%%%%%%%%%%%%%%%%%%%%%%%%%%%%%%%%%%%%%%%%%%%%
\begin{figure*}
\centering
\includegraphics[width = 0.7\textwidth]{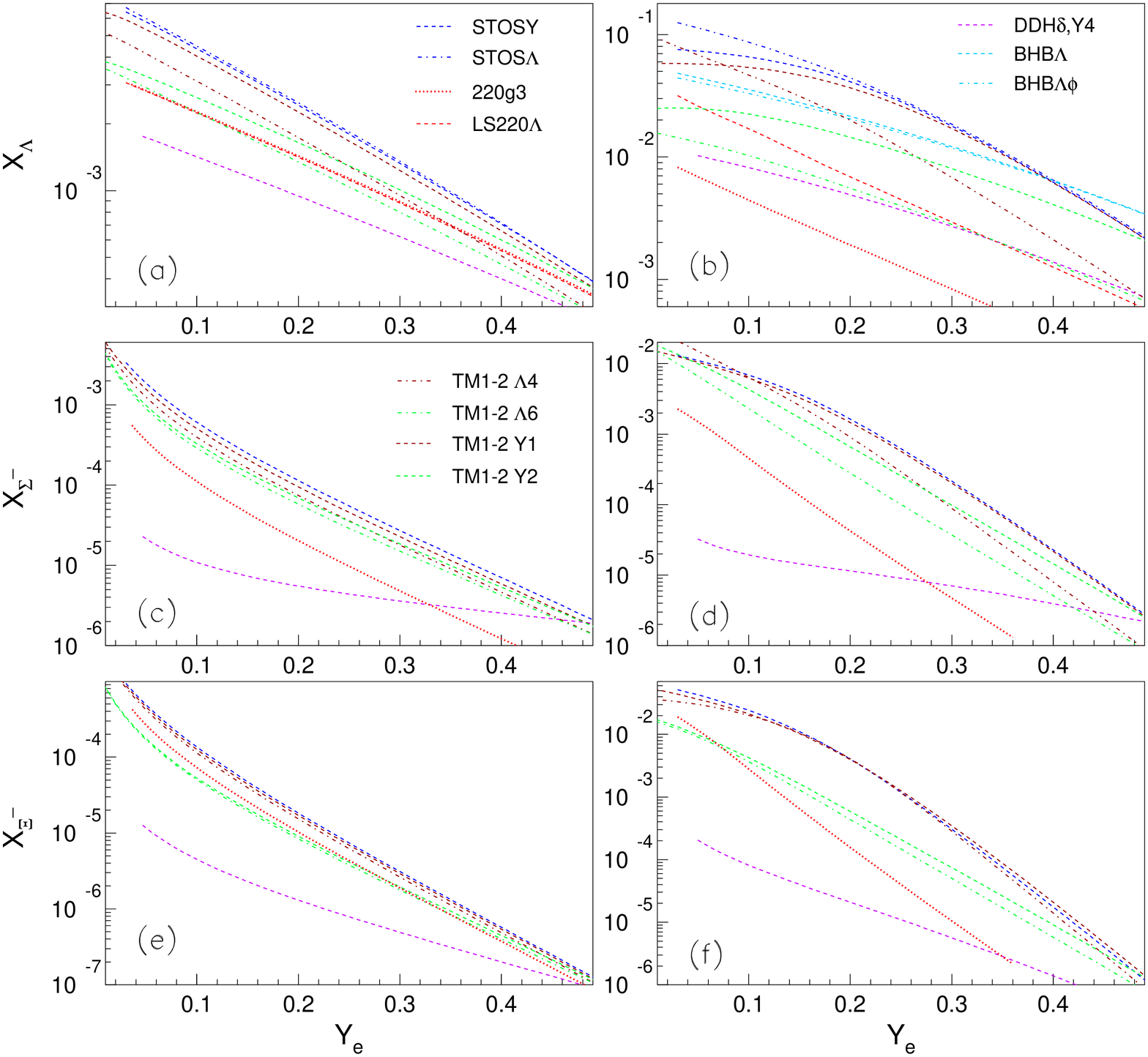}
\caption{(Color online) Fractions of different hyperons as a function of the
  electron fraction at a temperature of 25 MeV and for $n_B = 0.15\
  \mathrm{fm}^{-3}$ (left) and $n_B = 0.3\ \mathrm{fm}^{-3}$ (right)
  corresponding roughly to once and twice nuclear matter saturation
  density. The fractions of $\Sigma^{0,+}, \Xi^0$ are not shown since they are
  always much smaller. 
}
\label{fig:xiye}
\end{figure*}
%%%%%%%%%%%%%%%%%%%%%%%%%%%%%%%%%%%%%%%%%%%%%%%%%%%%%%%%%%%%%%%%%%%%%%%%%%%%%%%%
%%%%%%%%%%%%%%%%%%%%%%%%%%%%%%%%%%%%%%%%%%%%%%%%%%%%%%%%%%%%%%%%%%%%%%%%%%%%
\begin{figure*}
\centering
\includegraphics[width = 0.7\textwidth]{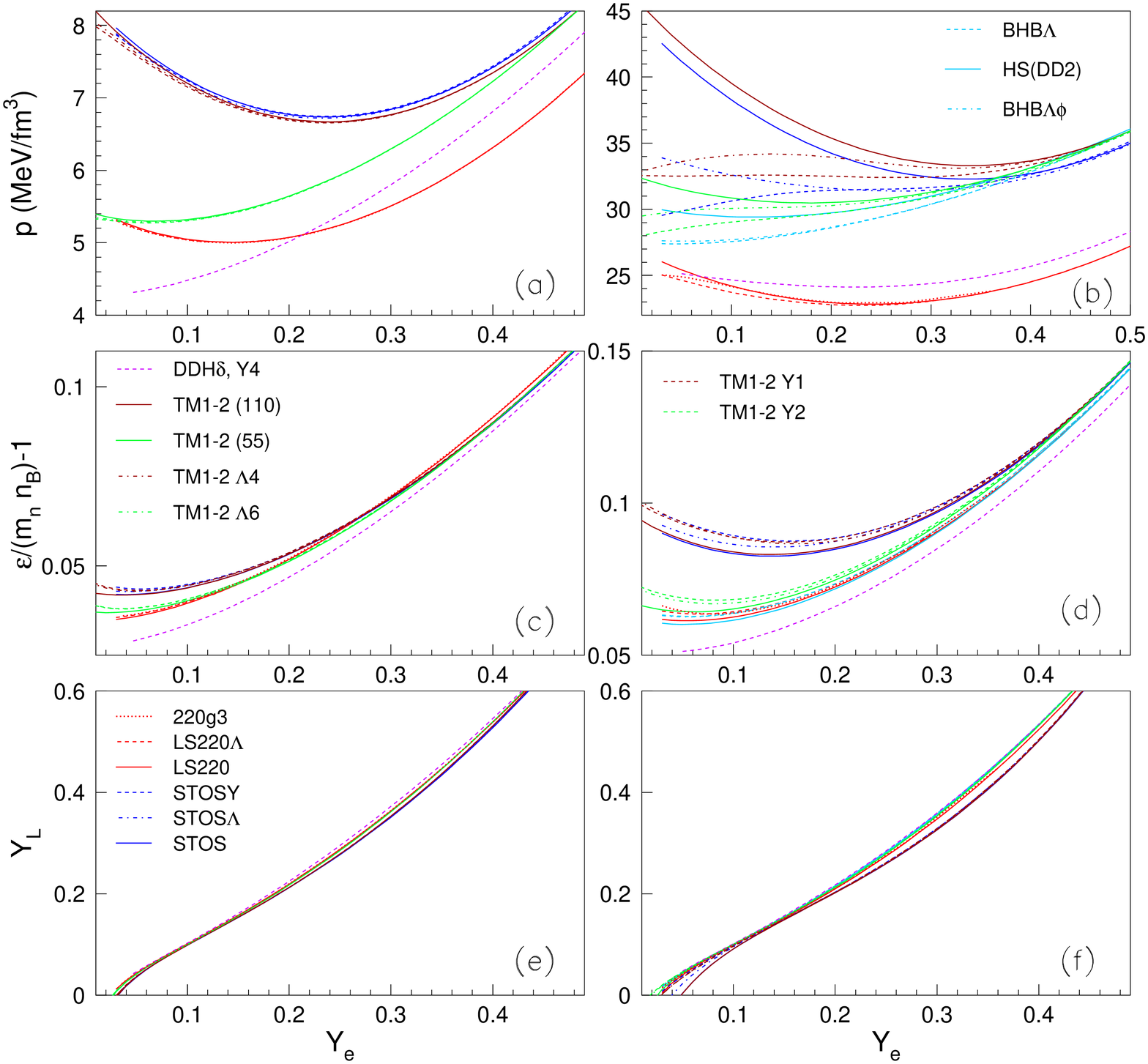}\hfill
\caption{(Color online) Thermodynamic quantities as functions of
  electron fraction $Y_e = Y_Q$ for a temperature $T = 25$ MeV and a
  baryon number density of $n_B = 0.15\ \mathrm{fm}^{-3}$ (left) and
  $n_B = 0.3\ \mathrm{fm}^{-3}$ (right). The upper panels show the
  pressure, the middle ones the internal energy per baryon with
  respect to the neutron mass and the lower ones the electron lepton fraction
  $Y_L = Y_e + Y_{\nu^e}$ under the assumption of
  $\beta$-equilibrium. No neutrinos are included in pressure and
  energy.}
\label{fig:thermofuncye}
\end{figure*}
%%%%%%%%%%%%%%%%%%%%%%%%%%%%%%%%%%%%%%%%%%%%%%%%%%%%%%%%%%%%%

%%%%%%%%%%%%%%%%%%%%%%%%%%%%%%%%%%%%%%%%%%%%%%%%%%%%%%%%%%%%%%%%%%%%%%%%%%%%%
\begin{figure*}
\centering
\includegraphics[width = .7\textwidth]{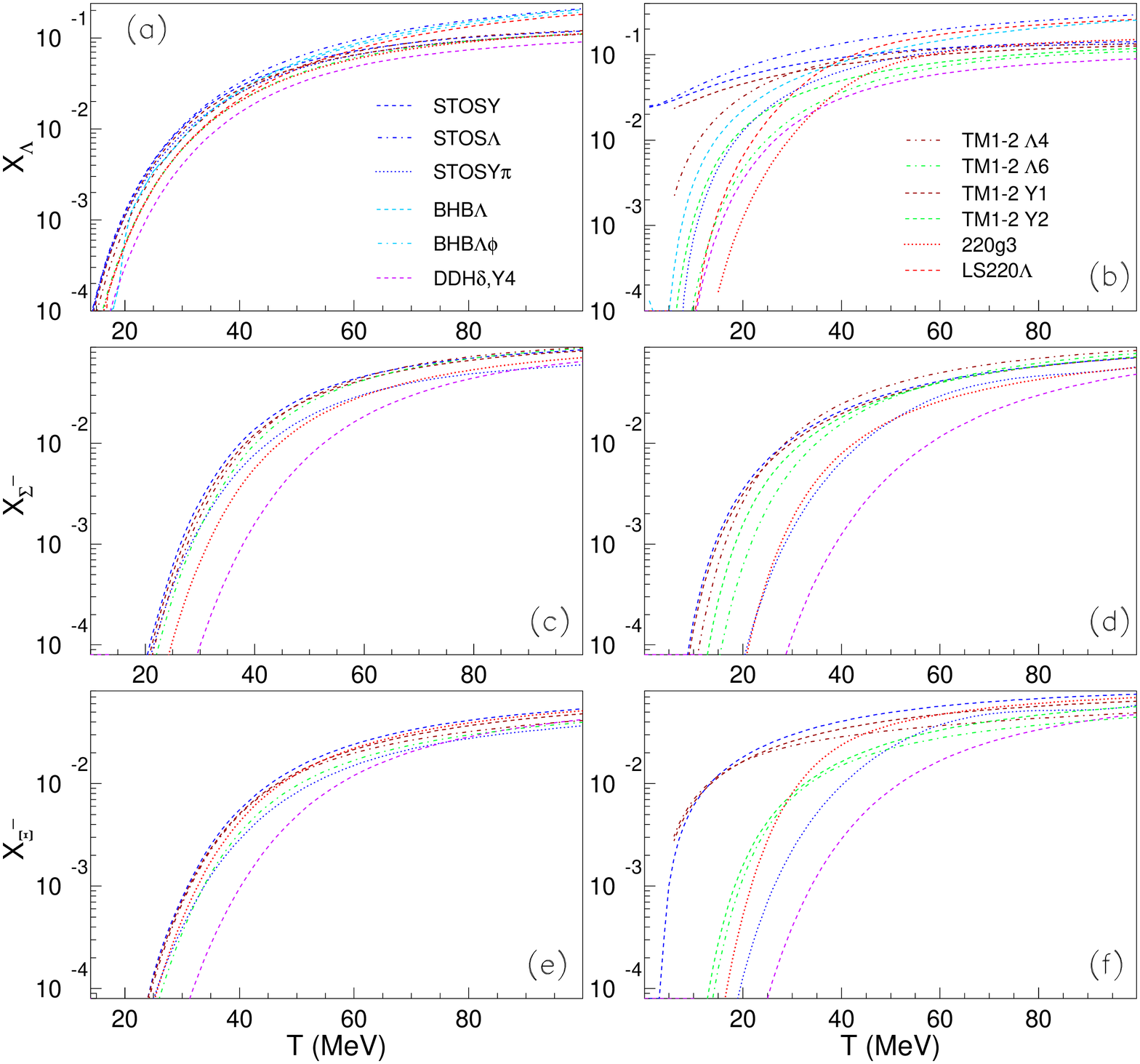}
\caption{(Color online) Same as Fig.~\ref{fig:xiye} but as a function of
  temperature for an 
  electron fraction $Y_e = 0.1$ and for $n_B
  = 0.15\ \mathrm{fm}^{-3}$ (left) and $n_B = 0.3\ \mathrm{fm}^{-3}$ (right).}
\label{fig:xifunctye10}
\end{figure*}
%%%%%%%%%%%%%%%%%%%%%%%%%%%%%%%%%%%%%%%%%%%%%%%%%%%%%%%%%%%%%%%%%%%%%%%%%%%%%%%%

As observed in Ref.~\cite{Fortin_14}, another point is that in many
models with hyperons compatible with the neutron star mass constraint,
see e.g.~\cite{Bednarek:2011gd,Weissenborn11c}, relatively large radii
of about 14 km for a non-rotating spherical neutron star with the
canonical mass of $1.4 M_\odot$ are obtained, well above some recently
suggested values, see e.g.~\cite{Ozel_15}. Radius determinations are
difficult and they are presently far from being as
reliable as the mass observations from
Refs.~\cite{Demorest_10,Antoniadis_13}.  The main problem is that the
extraction of radii from observations is much more model-dependent and
the low radii are not uncontested.  A summary and discussion of
different observational radius determinations can be found e.g. in
Ref.~\cite{Fortin_14}.  Anyway, concerning hyperons in neutron stars,
there are some examples with lower radii ~\cite{RikovskaStone:2006ta,Colucci2013,Banik2014,Oertel_14}.

Following the above remarks, we have chosen for our discussion of
thermal effects different models with very different behavior for cold
$\beta$-equilibrated neutron stars. As can be seen from of
Fig.~\ref{fig:eos3d}, where the mass-radius relations for a
non-rotating spherically symmetric neutron star are shown, the
obtained range in radii for intermediate masses is relatively
large. The onset of the additional degrees of freedom is clearly
visible in the different curves as an evident change in the slope. In
the models where only hyperons as non-nucleonic degrees of freedom are
present, the radius at a gravitational mass of $M = 1.4 M_\odot$ is
mainly determined by the nuclear part of the EoS, thus strongly
influenced by the parameter $L$, see Table~\ref{tab:nuclear}, since
hyperons appear only at densities above the central densities of these
stars. If pions are included, on the other hand, they show up already
at roughly saturation density and lower the radii already for neutron
stars with masses below $1.4 M_\odot$. A solution in order to obtain
lower radii could thus be that a mesonic contribution should not be
neglected. We should, however, be careful with a definite conclusion
here, since the models for including pions shown here are very crude,
see above, and the pion-nucleon interaction is neglected.

Depending on the hyperonic interaction chosen, the models give very
different maximum masses -- not all are compatible with the recent
constraints -- and the strangeness content, see the examples shown in
Fig.~\ref{fig:hypfractions}, covers a wide range, too. A summary of
the zero temperature results can be found in
Table~\ref{tab:nsresultsT0}. Concerning the EoS with non-nucleonic
degrees of freedom, which have been extended to finite temperature,
there are only a few which are compatible with the 2$M_\odot$
constraint, see e.g.~\cite{Dexheimer:2008ax}. In particular, the
existing EoS covering the whole range of temperature, electron
fraction and density relevant for core-collapse supernovae and binary
mergers either contain only $\Lambda$-hyperons~\cite{Banik2014,Gulminelli:2013qr,Peres_13,Shen:2011qu} or maximum cold neutron star masses well
below 2$M_\odot$ are obtained~\cite{Ishizuka_08}. Here we will show
first results extending the models of Ref.~\cite{Oertel_14},
compatible with existing constraints, to finite temperature and matter
not necessarily in $\beta$-equilibrium, i.e. for different electron
fractions, $Y_e = (n_{e^-} - n_{e^+})/n_B$.

As we will see, thermal effects favor the appearance of additional
particles and with increasing temperature the effect of the
interactions becomes less important. The reason is that the purely
kinetic thermal part dominates if the density does not become too
high. Let us stress that all the models we are showing here are based
on phenomenological models with parameters fixed to zero temperature
properties of nuclear matter, nuclei, nucleons and other hadrons. It
could be that these effective couplings depend on temperature,
although there are indications that this is not the case. In
Ref.~\cite{fedoseew_15} a RMF-model is compared with finite
temperature microscopic Dirac-Brueckner-Hartree-Fock calculations,
showing the thermal modifications of the effective couplings is almost
negligible. Similar conclusions are obtained with non-relativistic
phenomenological Skyrme type models~\cite{moustakidis_09,Fantina_private}.

In Fig.~\ref{fig:xiye} the fractions $X_i = n_i/n_B$ of $\Lambda$,
$\Sigma^-$ and $\Xi^-$ are shown as a function of the electron
fraction which equals the hadronic charge fraction for a constant
temperature of $T = 25$ MeV and two baryon number densities: $n_B =
0.15$ fm$^{-3}$ (left) and 0.3 fm$^{-3}$ (right). This value of the
temperature has been chosen since it corresponds in all present models
to an entropy per baryon $s_B$ between 1 and 3 $k_B$, 
values which are often
cited as typical conditions for proto-neutron stars, see
e.g.~\cite{Prakash:1996xs,Dexheimer:2008ax,Rabhi_11}. Other
hyperonic particle fractions are not shown since they are much
lower. The temperature is still low compared with the chemical
potentials, so that the abundances are dominated by the
latter. However, for the lower density, the temperature explains the
appearance of the hyperons, which at zero temperature would not be
present in matter. Therefore, the large negative charge chemical
potential in matter with low charge fractions favors negatively
charged particles resulting in very low abundances for the neutral and
positively charged hyperons, except for the $\Lambda$-hyperons. Here,
the lower mass compared with $\Sigma$-hyperons and Cascades
compensates the effect of the charge chemical potential.

All hyperonic fractions decrease with increasing $Y_Q$.  The reason is
that not only the charge chemical potential increases, but at the same
time the baryon number chemical potential decreases due to the fact
that the total baryon number remains constant: the proton fraction is
larger, and nuclear matter becomes more symmetric for $Y_Q$ close to
0.5, and, therefore the onset of hyperons is not favored. The effect
is more pronounced for the charged particles than for the neutral
$\Lambda$ due to the effect of $\mu_Q$. Qualitatively, all models show
a similar behavior. The quantitative differences due to the different
interactions are obviously more pronounced at higher density (right
panel), than at lower
density (left panel) where temperature is playing a larger role.  We
may also see how the hyperon fractions are sensitive to the density
dependence of the symmetry energy: a softer symmetry energy favor the
nucleonic degrees of freedom and the hyperon fractions are
smaller. This is clearly seen comparing TM1-2 Y1 and Y2, Y1 having a
harder symmetry energy and favoring larger fractions of hyperons. This
effect is more visible for $n_B=0.3$ fm$^{-3}$ (right panels), where
the temperature has a less important role.

No imprint of the strangeness-driven phase transition, see
Sec.~\ref{section:PT}, is visible in the presented curves for
LS$220\Lambda$ or TM1-2$\Lambda$4/6. The reason is that the critical
density lies above $n_B = 0.3$ fm$^{-3}$, the highest density shown
here.
\begin{figure}
\centering
\includegraphics[width = .9\columnwidth]{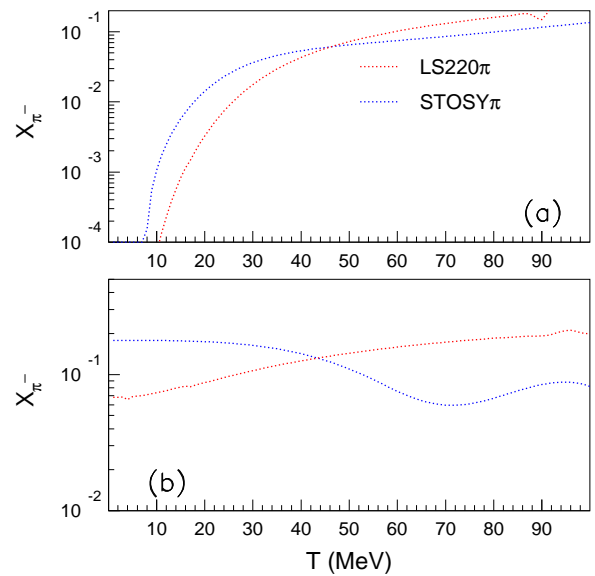}
\caption{(Color online) Same as Fig.~\ref{fig:xifunctye10}, showing  $X_{\pi^-}$ within the models where they are present at $n_B = 0.15$ fm$^{-3}$ (upper panel) and $n_B = 0.3$ fm$^{-3}$ (lower panel).}
\label{fig:xpifunctye10}
\end{figure}

In Fig.~\ref{fig:thermofuncye} the pressure,
the internal energy with respect to the neutron mass and the total
lepton fraction under the assumption of $\beta$-equilibrium are shown
within different EoS models. Comparing the purely nuclear models
(solid lines) with their counterparts containing additional particles,
it is evident that for given densities and a fixed temperature, the
pressure is lowered by the additional particles. As can be seen by
comparing STOS$\Lambda$ with STOSY, the more degrees of freedom
present, the lower the pressure. The two curves start to show a
difference upon onset of other hyperons than $\Lambda$-hyperons.

%%%%%%%%%%%%%%%%%%%%%%%%%%%%%%%%%%%%%%%%%%%%%%%%%%%%%%%%
\begin{figure*}
\centering
\includegraphics[width = 0.7\textwidth]{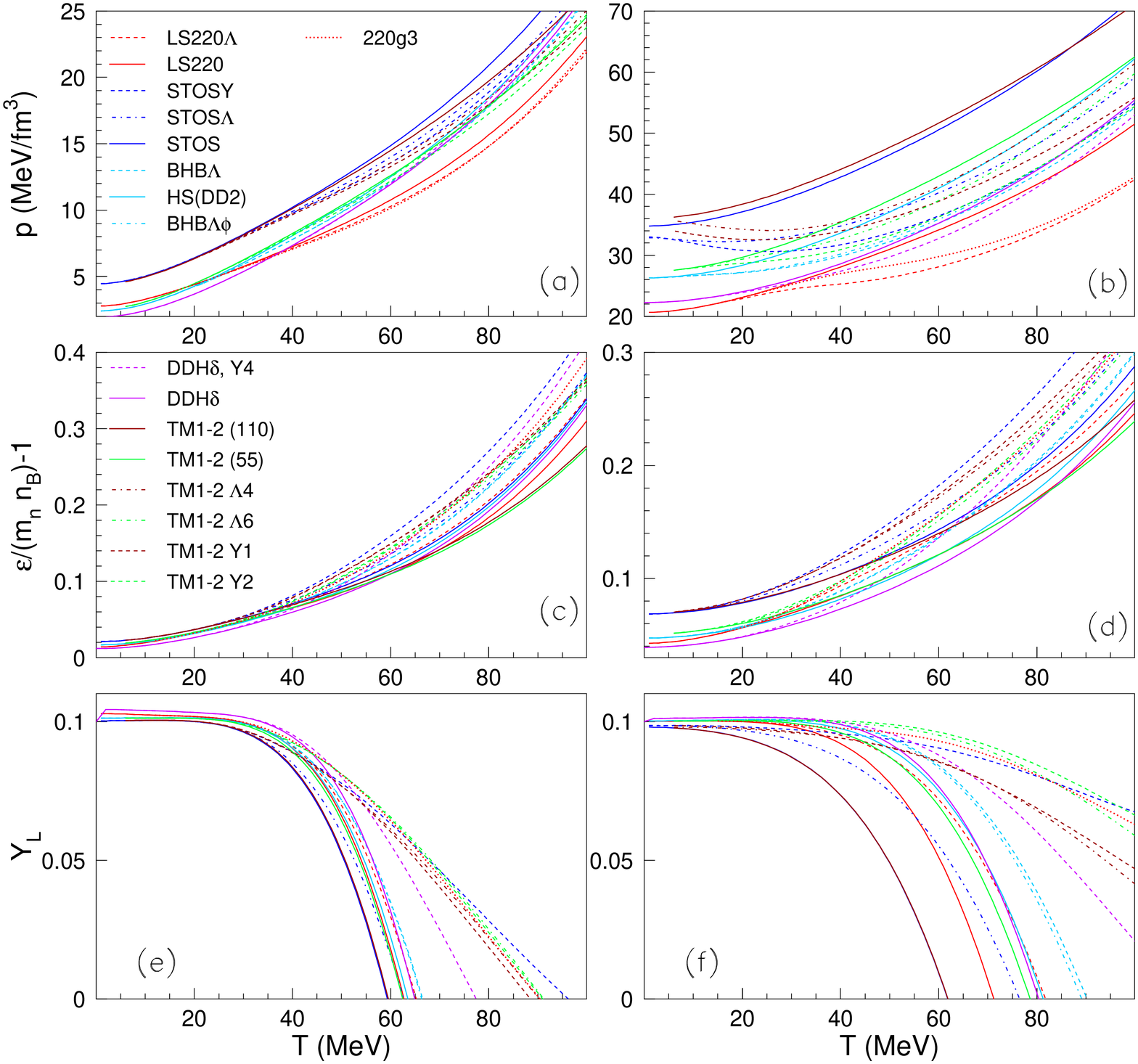}\hfill
\caption{
Same as Fig.~\ref{fig:thermofuncye} but as function of temperature 
for an electron fraction $Y_e = Y_Q = 0.1$. }

\label{fig:thermofunct}
\end{figure*}
%%%%%%%%%%%%%%%%%%%%%%%%%%%%%%%%%%%%%%%%%%%%%%%%%%%%%%%%%%%%
At a temperature of 25 MeV, the abundances of the additional particles
are still low, see above, such that the effect is not very
pronounced. The difference in pressure increases with decreasing $Y_e$
since the abundances increase. Already at saturation density, due to
thermal effects, there is a small reduction in pressure, but the
effect becomes obviously stronger with increasing density. At twice
nuclear matter saturation density there are still no hyperons in the
zero temperature EoS, but thermal effects induce a clear effect on the
EoS. Comparing the different hyperonic interaction models, in
particular the different versions based on TM1-2, it can be seen that
the interaction plays some role in determining the pressure.

In the middle panels, the internal energy defined as
\begin{equation}
\epsilon = \frac{\varepsilon}{n_B m_n} -1~,
\end{equation}
with $\varepsilon$ denoting the total energy density, is displayed.
In contrast to the pressure, which is decreased for given densities
and temperatures if hyperons are included, it is --slightly for the
conditions shown in Fig.~\ref{fig:thermofuncye}-- increased due to the
presence of hyperons.  This is understandable since hyperons are more
massive than nucleons, thus for a given baryon number density, if a
nucleon is replaced by a hyperon, the energy density is increased.

In the lower panels, the total electron lepton fraction is shown, $Y_L
= Y_e + Y_{\nu^e}$, assuming $\beta$-equilibrium with neutrinos. This
means that the neutrinos follow a Fermi-Dirac distribution with a
chemical potential equal to the electron lepton number chemical
potential. The latter can be calculated as
\begin{equation}
\mu_{\nu^e} = \mu_{L} = \mu_e + \mu_Q = \mu_e + \mu_p - \mu_n~,
\end{equation} 
with the electron chemical potential $\mu_e$ and the charge chemical
potential $\mu_Q$, given by the difference of proton and neutron
chemical potentials. The neutrino chemical potential is increasing
with increasing $Y_e$, such that in all models the neutrino fraction
increases with $Y_e$. At this temperature only very small differences
can be observed comparing the purely nucleonic models with the models
including non-nucleonic degrees of freedom. Only at very low $Y_e$, a
slightly higher $Y_L$ is obtained within the latter models.  $s_B$ of
1 $\sim$ 2 $k_B$ together with trapped neutrinos and a value $Y_L$ = 0.4 are
typical values inside a proto-neutron star, see for instance
Ref.~\cite{Prakash:1996xs}. In all the models shown here this would
correspond to charge fractions between 0.3 and 0.4 with no appreciable
difference between the nucleonic models and the others.

Depending on the progenitor, in the early post-bounce phase of a
core-collapse event, much higher temperatures can be reached, of up to
100 MeV, see e.g.~\cite{Peres_13,sumiyoshi_09,Char_15}. Therefore,
in Fig.~\ref{fig:xifunctye10}, the hyperonic particle fractions are
displayed as function of temperature, again for $n_B = 0.15$ (left)
and $n_B = 0.3$ fm$^{-3}$ (right) and in Fig.~\ref{fig:thermofunct} the
corresponding pressure, internal energy and $Y_L$. The electron
fraction has been fixed to a low value of $Y_e = 0.1$. As discussed
before, the hyperonic abundances become smaller for larger value of
$Y_e$ and typical values in the hot newly formed proto-neutron star
are more of the order 0.3, while the neutrino free mean path is small
and neutrinos are trapped inside the star.  We have nevertheless
chosen this low value to maximize the impact of non-nucleonic degrees
of freedom, it corresponds to a scenario when the star becomes
transparent to neutrinos.

In the upper panels of Fig.~\ref{fig:xifunctye10} the $\Lambda$-, in the
middle panels the $\Sigma^-$- and in the lowest panels the
$\Xi^-$-fractions are shown. They obviously increase all with
temperature and $X_\Lambda$ can reach more than 20 \% at 100 MeV even
at saturation density. At saturation density, all hyperons start to
show up at roughly 25 MeV, independently of the model, whereas at $n_B
= 0.3$ fm$^{-3}$, differences are visible and in some models,
$\Lambda$-hyperons exist already at very low temperatures.

The models allowing only $\Lambda$-hyperon states to be populated have
considerably higher $X_\Lambda$ than those including the whole baryon
octet. The reason is simply, since the baryon number density is fixed,
if other hyperons appear, they replace the $\Lambda$'s. Indeed, adding
up all the hyperon fractions, the total strangeness fraction is
similar within all models shown here. Again, as expected, the
interaction dependence is more pronounced at higher density. The
$\Xi^-$ and $\Sigma^-$ abundance in the DDH$\delta$-model is probably
much lower than in all other models due to the very strong
$YY$-repulsion in these channels for the given parameterization, see
Table~\ref{tab:hypcouplings}. Again, the role of the symmetry energy
becomes clear comparing the $\Sigma^-$ and $\Xi^-$ fractions of TM1-2
Y1 and Y2: for the harder symmetry energy the onset of hyperons occurs
at smaller temperatures.

In Fig.~\ref{fig:xpifunctye10}, the fractions of $\pi^-$ are shown
within the two models considered here which contain pions. $\pi^0$ and
$\pi^+$ have much lower abundances, the fractions stay below 1\%
except above 80 MeV and are therefore not shown. At $n_B = 0.15$
fm$^{-3}$, i.e. roughly saturation density, the $\pi^-$-fraction
exceeds 1\% at about 25 MeV and increases up to about 20\%. 
At twice this density, the fraction is less temperature dependent and remains
between 5 and 15\%. Despite this non-negligible abundances, the impact
on the EoS is less important than for hyperons, see below.

At which temperature do the non-nucleonic particles start to
considerably influence the EoS? In order to answer this question we
display in Fig.~\ref{fig:thermofunct} the pressure and the internal
energy as function of temperature for the same $Y_e$ and $n_B$ as
before. They clearly show the impact of the appearance of
non-nucleonic degrees of freedom. The softening due to the additional
degrees of freedom is again visible in the pressure and the internal
energy is increased upon their onset. Again, at $n_B = 0.15$
fm$^{-3}$, the particle content of the EoS has more influence on the
behavior of the thermodynamic quantities than the details of the
interaction, whereas, at twice this density, the different
parameterizations result in different values for pressure and internal
energy.

Concerning the lepton fraction assuming $\beta$-equilibrium displayed
in the lower panels, in the EoS with non-nucleonic degrees of freedom
systematically higher values are obtained. The reason is that the
presence of additional particles increases the charge chemical
potential. The strong drop in $Y_L$ occurs at higher temperatures,
too. The reason for this drop is that the neutrino chemical potential
is decreasing with increasing temperature in all models, much stronger
in the purely nucleonic models due to the lower $\mu_Q$ than in the
others. 

\begin{figure}
\includegraphics[width=0.95\columnwidth]{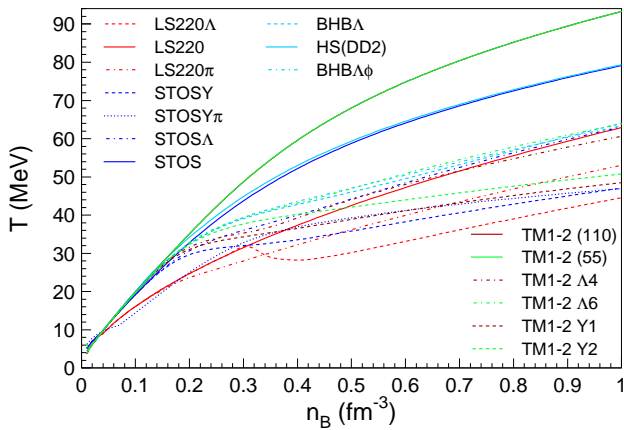} 
\caption{(Color online) Temperature as a function of baryon number density at constant entropy per baryon $s = 2 k_B$ and constant $Y_Q = 0.1$. Models including hyperonic degrees of freedom are compared with their purely nucleonic counterparts.  
\label{fig:entropy}}
\end{figure}

Another interesting point about including non-nucleonic degrees of
freedom is the fact that for a given entropy per baryon and electron
fraction the temperature is significantly lower within an EoS
including hyperons and/or pions than in a purely nuclear one, see
e.g.~\cite{Rabhi_11}. This can be observed from
Fig.~\ref{fig:entropy}, where for a fixed entropy per baryon, $s_B = 2
k_B$, and constant electron fraction $Y_e = Y_Q = 0.1$ the temperature
is shown as a function of baryon number density within different EoS
used within this paper. Upon the onset of the additional degrees of
freedom, i.e. about saturation density if pions are included and at a
slightly higher density if only hyperons are allowed, the temperature
curves considerably deviate from the purely nuclear EoS and increase
much less stringently with density. As pointed out before, a softer
symmetry energy gives rise to smaller hyperon fractions, and, as a
consequence the temperature increases faster with density for a fixed
entropy per baryon. The more degrees of freedom included, the lower
the temperature. This can be understood looking at the temperature of
a multicomponent Fermi gas at fixed entropy. For degenerate Fermi
particles, i.e. for low temperatures compared to the Fermi energies of
the system components, it is given by \cite{Steiner:2000bi}
\begin{equation}
T\sim\frac{s}{\pi^2} \left( \frac{\sum_i p_{F_i}^3}{\sum_i p_{F_i} \sqrt{p_{F_i}^2+\left(m_i^*\right)^2}} \right),
\end{equation}
where $p_{Fi}$ and $M^*_{Fi}$ are, respectively, the Fermi momentum
of component $i$ and corresponding effective mass, and $s$ is the
entropy per particle of the system.  It is obvious that, increasing the
number of degrees of freedom makes the temperature increase more
slowly because the Fermi momenta of the system components decrease.
This is a trivial thermodynamic effect: the appearance of hyperonic
species implies that the energy is shared among an increased number of
degrees of freedom, with consequently reduced thermal excitations for
each of them.

This qualitative result does not depend very much on the value of
$Y_e$ and is confirmed for trapped neutrinos,
too~\cite{Rabhi_11}. Assuming that in the region enclosed by the shock
at early post-bounce times, the entropy profile is only weakly
dependent on the EoS, this would mean that the non-nucleonic degrees
freedom considerably lower the temperature with obviously important
consequences for the neutrino distribution and the neutrino heating
mechanism.  The authors of Ref.~\cite{Mayle93} argue, however, that
the presence of negatively charged particles other than electrons
lowers the net electron number, releasing the electron degeneracy
energy and resulting finally in a higher temperature of the supernova
core.  Without performing realistic simulations, including neutrino
transport to determine the electron fraction of the EoS, this question
cannot be definitely answered.

%%%%%%%%%%%%%%%%%%%%%%%%%%%%%%%%%%%%%%%%%%%%%%%%%%%%%%%%
%%%%%%%%%%%%%%%%%%%%%%%%%%%%%%%%%%%%%%%%%%%%%%%%%%%%%%%%%%%%%%
%%%%%%%%%%%%%%%%%%%%%%%%%%%%%%%%%%%%%%%%%%%%%%%%%%%%%%%%%%%%
\begin{figure}
\centering \includegraphics[width =
  0.9\columnwidth]{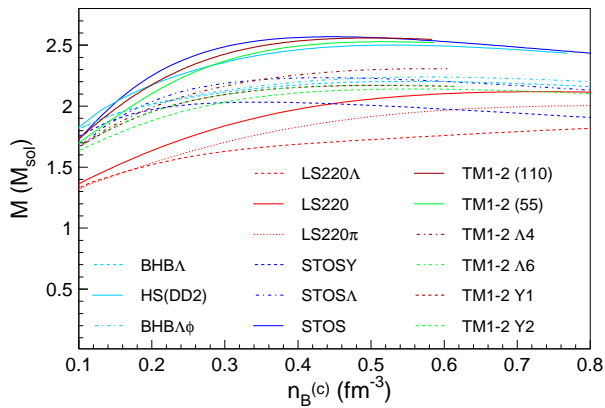}\hfill
\caption{Mass-central baryon density relations for isentropic spherical stars with $s = 4$ in neutrino-less $\beta$-equilibrium employing different EoS discussed in the text. }
\label{fig:mnbs4}
\end{figure}
%%%%%%%%%%%%%%%%%%%%%%%%%%%%%%%%%%%%%%%%%%%%%%%%%%%%%%%%%%%%%
Non-nucleonic degrees of freedom impact the maximum mass of a hot
proto-neutron star, too. In general, compared with the cold
$\beta$-equilibrated neutron star maximum mass, it is generally found
that at typical early proto-neutron star conditions of $s_B = 2 k_B$
and a fixed lepton fraction, $Y_{L} = 0.4$, the maximum mass is
increased for all EoS~\cite{Prakash:1996xs,Pons:1998mm}. Thermal
effects are, however, at this entropy value less important than the effect
of trapped neutrinos which induce a shift to higher $Y_Q$ with respect
to the cold neutrino-less $\beta$-equilibrium case. Indeed, inspecting
the upper panels of Figs.~\ref{fig:thermofuncye} and
\ref{fig:thermofunct}, the pressure is more strongly influenced by a
shift of $Y_e \sim 0.1$ to $\sim 0.3$ than between zero temperature
and roughly 25 MeV for all EoS shown there. An exception are the EoS
with a kaonic BEC considered in Ref.~\cite{Pons:2000xf,Pons:2000iy}. 

Since the abundances of hyperons and pions increase with decreasing
$Y_e = Y_Q$, during the deleptonisation of the hot proto-neutron star
the softening of the EoS with respect to the purely nucleonic one due
to the additional degrees of freedom becomes more pronounced, see
Fig.~\ref{fig:thermofuncye}. This could give rise to metastable
lepton-rich proto-neutron stars, see
e.g.~\cite{Prakash:1996xs,Pons:2000xf,chen_12,lugones_98,steiner_00,shao_11},
and Sec.~\ref{sec:bh_formation}, too.

%%%%%%%%%%%%%%%%%%%%%%%%%%%%%%%%%%%%%%%%%%%%%%%%%%%%%%%%%%%%%%%%
\begin{table}
\begin{center}
\begin{tabular}{l|cccc}
\hline
 Model& $M^{s=4}_{\mathit{max}}$& $f_S$& $n_B^{(c)}$ & $T_{\mathrm{max}}$ \\
&$M_\odot$ & & [fm$^{-3}$] & [MeV]\\ \hline \hline
DD2         &2.50& 0  &0.53 &118 \\ 
TM1 (STOS)  & 2.57& 0  & 0.45& 109 \\ 
TM1-2(L=111)  & 2.56& 0 &0.50 &104 \\ 
TM1-2 (L=55)  & 2.53& 0 & 0.52& 109 \\ 
LS220  &2.12 & 0 &0.74&123  \\ 
\hline
BHB$\Lambda$  &2.20&0.15 &0.56 & 105 \\ 
BHB$\Lambda\Phi$  & 2.24&0.14 & 0.57&104\\
STOS$\Lambda$  &2.23&0.13&0.43&72\\
STOSY  &2.06&0.15&0.41&75 \\
TM1-2  Y1 &2.17 &0.15&0.51&84\\
TM1-2  Y2 &2.14&0.16&0.56&89\\
TM1-2 $\Lambda$4& 2.31 &0.18&0.58&95 \\ 
TM1-2 $\Lambda$6& 2.30 &0.16&0.60&99 \\ 
LS220$\Lambda$  &1.85&0.15&1.06&122 \\ 
LS220$\pi$  &2.00&-&0.83&108 \\ 

\hline

\end{tabular}
\caption{ Results calculated within different models: maximum mass of
  an isentropic star with entropy per baryon $s_B = 4 k_B$ in
  $\beta$-equilibrium without neutrinos. Several quantities are listed
  for the maximum mass configuration: 
(a) The total strangeness
  fraction, $f_S$, representing the integral of the strangeness
  fraction $Y_s/3$ over the whole star as in
  Ref.~\cite{Weissenborn11c}, 
 (b)The central baryon number density
  (c) The highest temperature reached.
%\mo{\item If available, the time to BH collapse with a 40$M_\odot$ solar metallicity progenitor. }
In the upper part purely nucleonic models are listed and in the lower part those containing hyperons and/or pions are given. }
\label{tab:nsresults}
\end{center}
\end{table}

%%%%%%%%%%%%%%%%%%%%%%%%%%%%%%%%%%%%%%%%%%%%%%%%%%%%%%%%%%%%%%%%%%%%

\section{Impact of additional particles in astrophysical applications}
\label{sec:astro}
Since we are mainly interested in thermal effects on an EoS not
necessarily in $\beta$-equilibrium, we will, within this section, not
consider any impact of hyperons or pions on properties of old and cold
neutron stars, neither consider their cooling process. These points
will be discussed elsewhere within this volume. As already mentioned
earlier, the two systems predestinated to be influenced by the
presence of non-nucleonic degrees of freedom are compact star binary
mergers, neutron star-neutron star (NS-NS) and neutron star-black hole
(NS-BH) mergers as well as BH formation in a core collapse supernova
and the early post-bounce proto-neutron star evolution. Under some
conditions, in particular --as mentioned earlier-- a very soft EoS not
compatible with a 2 $M_\odot$ neutron star~\cite{sagert_12}, the
dynamics of an exploding supernova could be influenced, too: In
Ref.~\cite{sagert_09} it has been shown that a phase transition to
quark matter in the early post-bounce phase could induce a second shock
wave, helping the supernova to explode. In general, densities and
temperatures are, however, not high enough for non-nucleonic degrees
of freedom to be considerably populated and to have a noteworthy
impact on an exploding supernova. We will therefore not discuss this
scenario further here.

Since, despite recent efforts, only a few EoS with hyperons
and/or pions are presently available covering the full range of
thermodynamic variables necessary to perform realistic simulations of
the above cited events, this field is under current development and we
only want to give some hints on interesting results here.

\subsection{Binary mergers}
Coalescing relativistic binary systems containing compact objects,
either NSs or black holes, receive a great interest since they are important sources of gravitational waves (GW), potentially detectable with next generation detectors such as advanced LIGO/VIRGO or KAGRA. In addition, they are believed to produce short gamma-ray bursts and they may represent a major source of heavy r-process elements. For more details, see e.g. the reviews~\cite{Rosswog_15,Faber_12,Shibata_11}. 

The most promising track to obtain information on the EoS at high
densities, i.e. containing potentially non-nucleonic degrees of
freedom, is the post-merger phase. If the EoS supports the formation
of a hypermassive neutron star, the frequencies of its normal modes
are sensitive to the EoS and visible in the GW signal. The
measurements of their frequencies could tightly constrain NS masses
and radii since they are strongly correlated. It could even be
possible to give an estimate for the NS maximum mass
\cite{Sekiguchi_11,Bauswein_12,Bauswein:2013jpa,Bauswein:2014qla,bauswein15}.

Other possibilities to constrain the EoS are the GW signal from the
late inspiral, where the tidal deformation depends on the
EoS~\cite{Faber_12,Shibata_11,Read_13} or the sGRB rate which depends
on the maximum mass supported by the merger remnant~\cite{Fryer_15,Lawrence:2015oka}. r-process nucleosynthesis depends critically on
the EoS and matter composition \cite{Wanajo:2014wha,sekiguchi15}, too.

\subsection{Black hole formation}
\label{sec:bh_formation}
It is well established that a massive star, after having consumed all
its fuel, becomes gravitationally unstable, leading to a so-called
core-collapse event. The iron core of the progenitor star collapses,
bouncing back if the central density reaches roughly nuclear matter
saturation density due to the stiffening of the EoS related to nuclear
forces. A shock is formed, propagating outwards and at the center
remains a hot lepton rich proto-neutron star. The following evolution
is less well understood, see e.g.~\cite{Janka_06} for a detailed
review. If the ejected material is successfully unbound after bounce, a
neutron star is formed in a supernova explosion. If, however, the
expanding shock is not able to break through the infalling material,
the accretion pushes the proto-neutron star over its maximum mass
which subsequently collapses to a black hole (BH). 
Due to larger accretion rates, such failed
supernovae reach higher densities and temperatures than
their exploding counterparts, which makes them an interesting tool to
explore the EoS at high temperatures and supra-saturation densities.
Alternatively to this
BH formation scenario in a failed supernova during the first
second after bounce, a BH could equally well be formed at
later times in the so-called delayed formation process, either because
the proto-neutron star becomes unstable or because ejected material
falls back and causes the collapse to a black hole.

Independently of the detailed scenario, the time until BH formation
turns out to be sensitive to the underlying EoS. In spherical
simulations, see e.g.~\cite{pons_01,Nakazato10b}, the neutrino
emission is abruptly stopped upon BH formation with a clear imprint in
the observable neutrino signal. The time until BH formation, however,
not only depends on the EoS, but on other factors, too, such as for
instance the structure of the progenitor star or the rotation
rate~\cite{Sekiguchi:2010ja} such that the interpretation of such a
signal would not be free of ambiguities although it presents a
promising track to learn about the EoS of hot and dense matter. 

Let us start now the discussion with the delayed BH formation. We have
already mentioned that the hot proto-neutron star with trapped
neutrinos is lepton rich and that hyperons and/or pions are less
abundant than in cold neutron stars with very low $Y_e$. The same is
true for EoS with a kaon condensed phase or with a transition to quark
matter~\cite{Pons:1998mm,Pons:2000xf,Pons:2000iy,pons_01,lugones_98}.  Thus imagine that we start from a
proto-neutron star with few non-nucleonic degrees of freedom and a
baryon mass above the maximum mass of a cold neutron star with these
degrees of freedom. During cooling the proto-neutron star
deleptonizes, and more and more additional particle states are
populated until eventually the star exceeds its mass-limit and
collapses to a BH, see e.g.~\cite{pons_01,Keil95,Baumgarte96}. 

Recently, some simulations of BH formation in failed supernovae have
been performed employing EoS with pions and
hyperons~\cite{Peres_13,Ishizuka_08,sumiyoshi_09,Char_15,nakazato_12} or pions and quarks~\cite{Nakazato10a,nakazato13}. All simulations
show that the time until BH formation is sensitive to the underlying
EoS and that the softening due to the additional degrees of freedom,
be it hyperons, pions or quarks, considerably shortens the time until
collapse to a BH. Since additional particles appear only deep inside
the proto-neutron star, the neutrino signal is changed mainly by the
duration of the signal compared with a purely nuclear EoS, see
e.g. the detailed comparison in Ref.~\cite{Char_15}. This could be
different if the onset of the additional particles are accompanied by
a phase transition, be it hyperons as in the LS220$\Lambda$
EoS~\cite{Peres_13} or quarks~\cite{Nakazato10a,nakazato13}. A phase transition
sufficiently modifies the dynamics to induce an observable difference
in the neutrino signal.  In this context, Ref. \cite{Ohnishi_11} discusses
the possibility that the QCD critical point could be reached in BH
formation of core collapse.

An interesting correlation concerning the time until BH in a failed
supernova has been raised in Refs.~\cite{Hempel_11a,Steiner_12}. They
performed core-collapse simulations with a set of purely nuclear EoS
starting from a 40 $M_\odot$ progenitor with solar metallicity. They
observed that the maximum mass of an isentropic proto-neutron star at
$s_B = 4 k_B$ in neutrino-less $\beta$-equilibrium can be correlated
with the time until BH formation. In Fig.~\ref{fig:mnbs4} we display
the gravitational mass of such a proto-neutron star as function of the
central density for a set of different EoS discussed within this
paper. In Table~\ref{tab:nsresults} in addition to the corresponding
maximum mass, the central baryon number density, the highest
temperature reached and the integrated strangeness content of the
different models are given. It is obvious that the maximum mass varies
drastically between different EoS and shows in particular a strong
reduction for the hyperonic EoS with respect to the purely nuclear
ones. This can be understood since rather high temperatures are
reached and the strangeness content of these proto-neutron stars is
very high. Therefore, if the correlation with the time until BH
formation is confirmed within a larger set of different EoS containing
non-nucleonic degrees of freedom, too, then it could become a very
interesting tool to constrain thermal properties of the EoS.

\section{Summary and Conclusion}

Within this paper we have discussed the appearance of non-nucleonic
degrees of freedom in dense and hot matter relevant for the description
of neutron stars, compact binary mergers and core-collapse
supernovae. These non-nucleonic degrees of freedom could, thereby, be
hadronic, i.e. hyperons or mesons or nuclear resonances, or quarks. We
have put the emphasis on hyperons and mesons and discussed mainly two
aspects.

The opening of hyperonic degrees of freedom in dense matter could
happen smoothly or could be accompanied by a phase transition with a
considerable effect on the thermodynamics and the hydrodynamical
evolution of the system. We have presented here a complete study of
the low temperature phase diagram including the entire baryon octet
within the phenomenological non-relativistic Balberg and Gal model
\cite{Balberg97}.  We have shown that the different hyperonic thresholds
may be associated with thermodynamic instabilities, leading to first
order phase transitions.  These transitions can merge into a wide
coexistence zone if the production thresholds of different hyperonic
species are sufficiently close.  As a consequence, a huge part of the
phase diagram might correspond to phase coexistence between low-strangeness
and high-strangeness phases.  In contrast to the nuclear liquid-gas
phase transition which is strongly quenched, this result is only
slightly affected by Coulomb effects upon adding electrons and
positrons to fulfill the charge neutrality constraint. As well as the
phase transition to quark matter~\cite{sagert_09}, this phase
transition could affect the dynamics of core
collapse~\cite{Peres_13}. In addition, at finite temperature a
critical point associated with this strangeness-driven phase
transition shows up, which might have sizable implications for the
neutrino propagation in core-collapse supernovae.

There exist plenty of EoS models for cold neutron stars discussing the
composition in the core and the possible appearance of hyperons,
mesons, nuclear resonances or quarks. This does not hold for matter at
finite temperature, where much less models exist, in particular if EoS
models are excluded whose cold $\beta$-equilibrated version is not
stiff enough to reproduce a 2 $M_\odot$ neutron star. However, even if
it turns out that finally, all these components are absent from cold
neutron stars, they could be populated at finite temperature due to
thermal effects. Within this paper, we have discussed results for
several phenomenological EoS models containing hyperons and/or pions
at finite temperature and for various hadronic charge fractions $Y_Q$. 

The models behave very differently concerning the cold neutron star
EoS. They cover a large range of neutron star radii, maximum masses --
not all of them are compatible with a maximum mass of $2 M_\odot$ --
and strangeness content. In particular, we have extended some recent
RMF parameterizations from Ref.~\cite{Oertel_14} to finite temperature
which contain the complete baryon octet and which give radii of the
order 12-13 km for a star with $M = 1.4 M_\odot$, maximum masses above
2 $M_\odot$ and a relatively important strangeness content. 

Qualitatively similar results are obtained within all EoS models,
independently of the underlying interaction. Major outcomes are that
hyperons and/or pions can become abundant already at saturation
density if the temperature exceeds roughly 25 MeV. Their abundance
increases with decreasing $Y_Q$, too, leading to the possibility of
meta-stable hot proto-neutron stars and a delayed collapse to a black
hole, see e.g.~\cite{pons_01}. Let us emphasize that a more
quantitative analysis of these results allowing to clarify the
composition and thermodynamic properties of hot and dense baryonic
matter would need additional constraints from future experimental data
on hyperonic interactions and/or ab-initio calculations of baryonic
matter with hyperons and mesons.

\section*{Acknowledgments}
This work has been partially funded by the SN2NS project 
ANR-10-BLAN-0503 
and by Project PEst-OE/FIS/UI0405/2014 
developed under the initiative QREN financed by the UE/FEDER through the
program COMPETE/FCT,
 and it has been supported by
NewCompstar, COST Action MP1304.
Ad. R. R acknowledges partial support from the Romanian National
Authority for Scientific Research under grants 
PN-II-ID-PCE-2011-3-0092 and PN 09 37 01 05
and kind hospitality from LPC-Caen and LUTH-Meudon.

% BibTeX users please use
 \bibliographystyle{epj}
 \bibliography{epja}
%

\end{document}